\documentclass[11pt,a4paper]{article}
\usepackage[T1]{fontenc}
\usepackage{lmodern}
\usepackage{jheppub}
\usepackage{amsthm,mathtools,microtype}
\let\originaltableofcontents\tableofcontents
\renewcommand{\tableofcontents}{\begingroup\small\renewcommand{\baselinestretch}{1.0}\selectfont\originaltableofcontents\endgroup}
\renewcommand{\afterTocRuleSpace}{\clearpage}
\newtheorem{theorem}{Theorem}[section]
\newtheorem{lemma}[theorem]{Lemma}
\newtheorem{corollary}[theorem]{Corollary}
\theoremstyle{remark}

\hypersetup{pdfauthor={Yi Liang},pdftitle={Local failure of disjoint additivity}}
\title{Local failure of disjoint additivity}
\author{Yi Liang\textsuperscript{*}\note[*]{ORCID: \href{https://orcid.org/0009-0006-3168-972X}{0009-0006-3168-972X}.}}
\affiliation{Independent Researcher, Beijing, China}
\emailAdd{iantrans2042@gmail.com}
\abstract{Local Gauss constraints can prevent commuting regional algebras from generating the full algebra assigned to their union. We investigate disjoint additivity on the full magnetic physical subspace of finite-group quantum-double models while keeping the original lattice edge supports fixed. Every finite non-Abelian group exhibits a counterexample on vertex-disjoint regions contained within a contractible disk on a regular square torus. For a fixed eight-edge loop and a specified peripheral contact family, additivity fails if and only if the group is non-Abelian and the selected contacts cover all four interior faces of the loop. Leaving an inner face uncovered restores additivity; for the same loop, so does any second region avoiding the specified testing neighborhood. A peripheral projector restricts the exterior tests on a loop transition, allowing their product to preserve the constraints even when the loop factor does not. We construct union observables that deterministically separate orthogonal relative-sign superpositions that yield identical outcome statistics under every measurement whose effects lie in the generated regional algebra. For a separate wrapping pair, we also determine the complete matrix-block inclusion, identifying missing center-orbit coherences, and construct an independent regular-torus counterexample for the eight-dimensional Kac--Paljutkin Hopf algebra.
}
\begin{document}
\maketitle
\section{Introduction}
\label{sec:introduction}

Describing spatial locality in a quantum system requires assigning observables to regions and specifying how those assignments compose, in keeping with the observable-first framework of algebraic quantum field theory~\cite{HaagKastler1964}. In unconstrained finite-dimensional spin systems, the spatial tensor-product decomposition of the state space provides both ingredients directly. Gauge constraints, however, can prevent the physical Hilbert space from inheriting that decomposition: in $(2+1)$-dimensional $\mathbb Z_2$ lattice gauge theory, closed electric strings crossing a cut obstruct a direct product of physical regional spaces, motivating an early construction that admits strings opening on the boundary~\cite{BuividovichPolikarpov2008ElectricStrings}. Algebraic and extended-Hilbert-space approaches adopt differing prescriptions for regional observables and traces~\cite{CasiniHuertaRosabal2014GaugeEntropy,Donnelly2012GaugeEntropy}. Consequently, a kinematic tensor product by itself settles neither the definition of physical regional algebras nor how they compose. Finite-dimensional quantum-double and string-net lattice models offer concrete settings where the interplay among local constraints, topological structure, and spatial support can be analyzed exactly~\cite{Kitaev2003FaultTolerantAnyons,LevinWen2005StringNet}.

Once a regional-algebra assignment is specified, a central composition problem is whether the algebras attached separately to two disjoint regions generate the full algebra of their union. Mutual commutativity does not settle this: commutativity concerns the relative ordering of operations from distinct regions, whereas disjoint additivity requires their generated algebra to exhaust the union algebra. For unconstrained spins, products of regional operators span the union algebra regardless of the state. The Bell states $(|00\rangle\pm|11\rangle)/\sqrt2$ have identical single-spin marginals, yet the product $X\otimes X$ of the two local Pauli spin flips has opposite eigenvalues on them. In our magnetic examples, by contrast, two orthogonal superpositions with opposite relative signs give identical statistics for every observable generated by the two regional algebras, yet an observable supported on the union distinguishes them with certainty. This algebraic generation question is distinct from defining subregion entanglement entropies or choosing boundary centers~\cite{Donnelly2014NonabelianEntropy,GhoshSoniTrivedi2015GaugeEntropy,AokiEtAl2015GaugeEntropy,SoniTrivedi2016GaugeEntropyAspects}. Here we assign to each region the physical $C^*$-algebra generated by the restrictions of kinematic operators supported on its original edges that map the full magnetic physical subspace into itself.

A foundational program in algebraic quantum field theory connects local observables to superselection sectors and particle statistics~\cite{DoplicherHaagRoberts1971,DoplicherHaagRoberts1974}. Extended operators connect generalized symmetries to additivity and Haag duality for regional algebras~\cite{CasiniEtAl2021EntropicOrder,CasiniMagan2021Completeness,GKSW2015}. Harlow et al.\ proposed disjoint additivity as a locality condition that can survive the failure of ordinary additivity. In their factorizable compact Lie group setting, averaging gives invariant operator representatives without enlarging support; suitable commuting local subgroups then yield disjoint additivity for nonadjacent regions~\cite{HarlowShaoSorceSrivastava2025Disjoint}. Violations constructed by projecting onto global symmetry sectors or prescribing extended-operator content~\cite{ShaoSorceSrivastava2025,JiaTian2026SymTFTAlgebras} motivate examining a complementary lattice setting in which local magnetic constraints are enforced while keeping the entire physical space and the original edge supports intact.

Holfester and Sorce examined this setting for non-invertible Gauss laws \cite{HolfesterSorce2026Gauss}. Proposition~4.3 in version~2 of their work states that magnetic group-double models satisfy disjoint additivity whenever the two edge regions are vertex-disjoint. We use the same magnetic physical space and the same edge supports, but also close the regional operator images under adjoints on the physical space. The union witness itself has a preserving representative on its prescribed support; its physical restriction lies outside even the enlarged regional join, and hence outside the algebra generated by the original images. A product can preserve the constraints even when one factor does not. Edge-wise inversion relates the two magnetic orientation conventions without changing supports (Section~\ref{subsec:magnetic-space}).

\begin{figure}[!htbp]
\centering
\includegraphics[width=\textwidth]{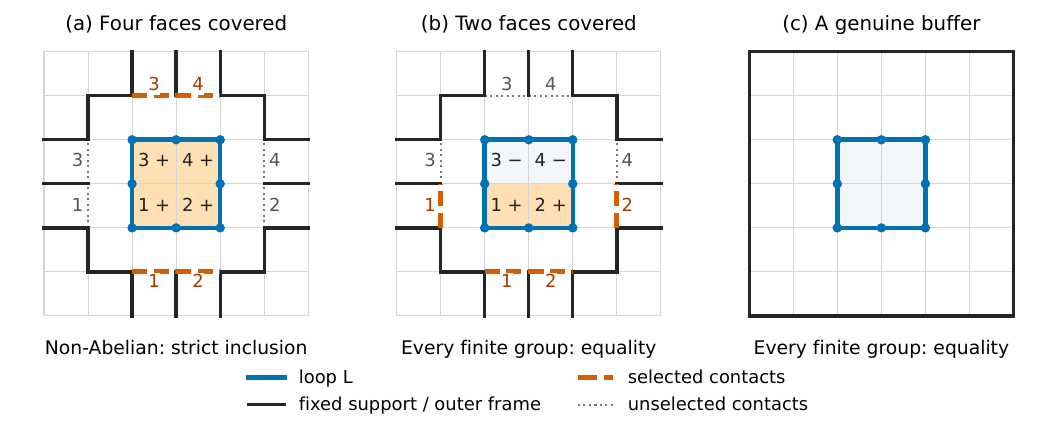}
\caption{In (a,b) each 24-edge peripheral support is at vertex-set distance one from the blue eight-edge loop $L$, yet the two layouts give different answers for non-Abelian groups. All supports lie in a common contractible disk of a square torus with at least eight vertices in each periodic direction. In both panels four selected contacts (dashed orange) join a common 20-edge base (black). Numerals link each inner face to its two possible exterior contacts; shading and $+$ mark covered faces, $-$ uncovered ones. Dotted unselected contacts and other pale edges belong to neither support, but their labels still enter the magnetic vertex constraints. (a) All four faces are covered: additivity fails for every finite non-Abelian group. (b) Only two are covered: additivity holds for every finite group. (c) The black 24-edge outer frame is not in the contact family. At vertex-set distance two from $L$, it avoids the boundaries of the four inner and eight adjacent outer faces; additivity holds for every finite group (Section~\ref{sec:buffer-restoration}).}
\label{fig:local-criterion}
\end{figure}

Our main obstruction is local in the ambient lattice geometry. For every finite non-Abelian group, we exhibit two vertex-disjoint edge supports contained in a single contractible disk of a regular square torus whose physical algebras commute yet fail to generate the physical algebra of their union. Figure~\ref{fig:local-criterion}(a,b) compares two 24-edge peripheral supports at vertex-set distance one from the same eight-edge square loop. They differ only in which four of eight possible contact edges are selected. Each inner face is paired with two exterior contacts; selecting either covers that face. For any subset of the eight contacts added to the same peripheral base, disjoint additivity fails precisely when the group is non-Abelian and all four inner faces are covered. If even one face is uncovered, the regional algebras generate the full union algebra. For the same loop, additivity also holds for any second region that avoids the boundary edges of the four inner and eight adjacent outer faces; in particular, a vertex-set graph distance of at least two from the loop suffices for any finite group.

{\samepage The underlying mechanism rests on how regional operators are tested against exterior configurations. An operator supported on the loop preserves the physical subspace only if it respects the vertex constraints for every valid exterior completion of its input. At each side midpoint, loop and exterior edges alternate around the vertex, forming a cusp. In~the permutation group $S_3$, let $e$ be the identity and $t=(12)$. One physical completion of the uniform loop $e$ puts $x$ on the vertical column through the lower cusp and $e$ elsewhere. Changing the loop to $t$ with the exterior fixed requires $txt^{-1}=x$ at this vertex. For $x=(123)$ the left side is $(132)$; $x=e$ passes. With all four faces covered, project the peripheral edges onto identity labels and exchange the two uniform loop configurations. On every physical basis configuration that survives both factors, the four inward cusp labels are $e$, so the exchange respects the constraints. The product annihilates the remaining basis configurations without changing the physical space. Using center-orbit projectors, we show that this composite observable lies outside the entire star algebra generated by the two regional algebras, not merely outside a single factorized presentation. Our restoration proofs identify the local exterior variations and supported unitary transports that reinstate the unrestricted completion test.\par}

Two additional constructions extend this analysis in complementary directions. First, for a separate pair of wrapping regions, we determine the full matrix-block inclusion: the missing coherences connect distinct orbits of the group center at fixed connecting-edge labels. Second, we construct an independent counterexample for the eight-dimensional Kac--Paljutkin Hopf algebra $H_8$~\cite{KacPaljutkin1965EightDimensional}, replacing diagonal configuration constraints with invariant-tensor relations. Because an abstract equivalence between lattice models need not preserve the regional subalgebras assigned to fixed edge sets \cite{BuerschaperEtAl2013Hierarchy,BuerschaperEtAl2013EMDuality,YanChenCui2022HopfRibbons}, our proof tracks the full physical-subspace image, the prescribed edge supports, and the physical adjoints directly on the original lattice.

Throughout this paper, we work on the full magnetic physical subspace, imposing neither electric plaquette projectors nor restrictions to fixed global-flux sectors, and making no restriction to the ground space of any specific Hamiltonian. The remainder of the paper is organized as follows. Section~\ref{sec:regional-algebras} formulates the regional observable prescription and the exterior-completion criteria. Sections~\ref{sec:local-criterion} and~\ref{sec:buffer-restoration} establish the local contact obstruction and the restoration of additivity outside the testing neighborhood. Sections~\ref{sec:wrapping-classification} and~\ref{sec:hopf-counterexample} present the wrapping-region block classification and the Kac--Paljutkin Hopf-algebra counterexample, and Section~\ref{sec:discussion} compares these results with alternative regional-composition frameworks.

\section{Magnetic constraints and regional observable algebras}
\label{sec:regional-algebras}

Formulating a regional locality statement requires specifying both a physical state space and a rule that assigns observable algebras to regions. Different assignments arise from introducing boundary extensions~\cite{DonnellyFreidel2016LocalSubsystems} or from selecting specific gauge-invariant subalgebras and centers~\cite{CasiniHuertaRosabal2014GaugeEntropy}. In Abelian lattice gauge theory, for instance, moving between electric- and magnetic-center prescriptions trades boundary electric generators for magnetic ones with which they need not commute, changing both the regional algebras and the operators omitted when those algebras are combined~\cite{Radicevic2014AbelianEntanglement}. Here we instead hold the original edge tensor factors and the full magnetic physical space fixed, and ask whether the algebras assigned to two vertex-disjoint regions generate the algebra assigned to their union.

\subsection{The physical space and lattice conventions}
\label{subsec:magnetic-space}

We consider a finite square lattice embedded in a torus, with edge Hilbert spaces $\mathcal H_e$ and full kinematic state space $\mathcal H=\bigotimes_e\mathcal H_e$. In the dual-frame formulation of quantum-double models, magnetic flatness constraints act at the lattice vertices~\cite{Kitaev2003FaultTolerantAnyons,HolfesterSorce2026Gauss}. Let $P_v$ denote the mutually commuting orthogonal projectors enforcing these vertex conditions. The joint physical projector and physical subspace are
\begin{equation}
 P=\prod_v P_v,\qquad \mathcal H_0=\operatorname{ran}P.
 \label{eq:physical-space}
\end{equation}
We impose solely the magnetic vertex projectors $P_v$, without restricting $\mathcal H_0$ to a fixed topological flux sector and without applying electric plaquette projectors. The explicit group-theoretic and Hopf-algebraic realizations of $P_v$ are defined in the sections that follow.

In the group model, each edge carries the group algebra $\mathbb C[G]$ of a finite group $G$ with identity $e$. We index the vertices of an $N\times M$ torus by $(i,j)\in\mathbb Z_N\times\mathbb Z_M$. Horizontal edges $h_{i,j}$ are oriented eastward from $(i,j)$ to $(i+1,j)$, and vertical edges $v_{i,j}$ are oriented northward from $(i,j)$ to $(i,j+1)$. In each group-basis configuration, the symbols $h_{i,j}$ and $v_{i,j}$ denote the group elements assigned to the corresponding oriented edges. The counterclockwise magnetic constraint at vertex $(i,j)$ and the total physical projector enforcing this condition at all vertices are given by
\begin{equation}
 C_{i,j}=h_{i,j}v_{i,j}h_{i-1,j}^{-1}v_{i,j-1}^{-1}=e,
 \qquad P=\prod_{i,j}\mathbf1[C_{i,j}=e].
 \label{eq:local-magnetic}
\end{equation}
An edge region $R$ is a subset of lattice edges, and we write $V(R)$ for the set of their endpoints. Two regions $R_1$ and $R_2$ are vertex-disjoint if $V(R_1)\cap V(R_2)=\varnothing$. For nonempty regions their vertex-set distance is the minimum graph distance between an endpoint of $R_1$ and an endpoint of $R_2$. All geometries below are formulated on a regular square torus; explicit dimensions and support configurations are specified in each setting. The non-cocommutative Hopf construction in Section~\ref{sec:hopf-counterexample} introduces its own edge orientations and invariant-tensor conventions.

The counterclockwise convention in~\eqref{eq:local-magnetic} is unitarily equivalent to the clockwise convention of Ref.~\cite{HolfesterSorce2026Gauss} by an on-site transformation that preserves edge supports. Write $a=h_{i,j}$, $b=v_{i,j}$, $c=h_{i-1,j}$, and $d=v_{i,j-1}$. Reading the incident edges counterclockwise from the east edge gives $C(a,b,c,d)=abc^{-1}d^{-1}$, whereas the source's clockwise word is $S(a,b,c,d)=ad^{-1}c^{-1}b$. The edge-inversion unitary $U_{\mathrm{inv}}=\bigotimes_\epsilon U_\epsilon$, with $U_\epsilon|g\rangle=|g^{-1}\rangle$, obeys $S(a^{-1},b^{-1},c^{-1},d^{-1})=a^{-1}C(a,b,c,d)^{-1}a$. Hence $U_{\mathrm{inv}}PU_{\mathrm{inv}}^\dagger=P_{\mathrm{cw}}$, where $P_{\mathrm{cw}}$ is the clockwise magnetic projector. Its tensor-product form transports supported preserving lifts, physical adjoints, and regional joins between the two magnetic physical spaces. This comparison does not assert equivalence of the electric terms of a full Hamiltonian.

\subsection{Supported physical operators}
\label{subsec:supported-operators}

For an edge region $R$, we define the regional Hilbert space $\mathcal H_R=\bigotimes_{\epsilon\in R}\mathcal H_\epsilon$, with $R^c$ denoting the complementary set of edges. Write $\mathcal B(\mathcal H_R)$ for the algebra of all linear operators on $\mathcal H_R$. The kinematic lift of $A_R\in\mathcal B(\mathcal H_R)$ to $\mathcal H=\mathcal H_R\otimes\mathcal H_{R^c}$ is $\widehat A_R=A_R\otimes I_{R^c}$, with $I_{R^c}$ the identity on $\mathcal H_{R^c}$. The set of kinematic lifts supported on $R$ that leave the physical subspace $\mathcal H_0$ invariant, together with their restrictions to $\mathcal H_0$, are defined by
\begin{align}
 K_R&=\{\widehat A_R:(1-P)\widehat A_RP=0\},\nonumber\\
 \mathcal A_0(R)&=\{\widehat A_R\vert_{\mathcal H_0}:
                          \widehat A_R\in K_R\}.
 \label{eq:raw-physical-image}
\end{align}
Preservation is required on all of $\mathcal H_0$, with the same regional operator acting for every physical exterior completion. This one-way condition does not require $[\widehat A_R,P]=0$, which would make the lift reducing. The raw physical image $\mathcal A_0(R)$ is a unital associative algebra, but it need not be closed under physical adjoints: for $T\in\mathcal A_0(R)$, the adjoint $T^\dagger$ on $\mathcal H_0$ may have no preserving lift supported on $R$. We define the regional observable algebra as the generated finite-dimensional $C^*$-algebra
\begin{equation}
 \mathcal A(R)=C^*\bigl(\mathcal A_0(R)\bigr),
 \label{eq:regional-star-algebra}
\end{equation}
with adjoints taken on $\mathcal H_0$ and without assuming that every element has a supported lift in $K_R$. This is an algebraic enlargement, not a claim of supported implementability. Separating a union-supported preserving operator from the enlarged regional join therefore also separates it from the algebra generated by the raw regional images.

Let $\mathcal A(R_1)\vee\mathcal A(R_2)$ denote the $C^*$-algebra on $\mathcal H_0$ generated by the two regional algebras. Lifts supported on either region are also supported on their union, so
\begin{equation}
 \mathcal A(R_1)\vee\mathcal A(R_2)
 \subseteq \mathcal A(R_1\cup R_2).
 \label{eq:additivity-inclusion}
\end{equation}
Disjoint additivity asserts equality in~\eqref{eq:additivity-inclusion} for vertex-disjoint regions. The group examples below give strict inclusion even when the regional algebras commute. For the wrapping group pair, all three raw images $\mathcal A_0(R_1)$, $\mathcal A_0(R_2)$, and $\mathcal A_0(R_1\cup R_2)$ are star-closed, so that example also violates additivity under the one-way-preservation definition of Ref.~\cite{HolfesterSorce2026Gauss}. This star-closure statement concerns that specific pair, not arbitrary group regions. In the non-cocommutative Hopf construction, the proof excludes generation in $\mathcal A(R_1)\vee\mathcal A(R_2)$ directly, without assuming that $\mathcal A_0(R_2)$ is star-closed. The separating union-supported operator $W$ in each counterexample satisfies $[W,P]=0$. Its physical restriction therefore belongs to the union algebra of reducing lifts, whereas the join generated by reducing regional lifts is contained in $\mathcal A(R_1)\vee\mathcal A(R_2)$; the same obstruction applies to that more restrictive prescription.

\subsection{Exterior fibres and star closure}
\label{subsec:fibre-criterion}

For constraints diagonal in a product basis, preservation can be checked entry by entry. Fibre inclusion gives the standard preorder pattern of a structural matrix algebra~\cite{GogicTomasevic2025Structural,DascalescuIovanovPredut2013Structural}. We use this description to track all physical exterior completions. It is not applied to the non-diagonal Hopf construction of Section~\ref{sec:hopf-counterexample}.

Fix a region $R$, and let $r$ and $x$ index orthonormal basis states for $\mathcal H_R$ and $\mathcal H_{R^c}$, respectively. Let $\Omega$ denote the set of physically allowed joint basis configurations satisfying all vertex constraints. The diagonal projector onto $\mathcal H_0$ and the allowed exterior fibre of a regional configuration $r$ are given by
\begin{equation}
 P=\sum_{(r,x)\in\Omega}|r,x\rangle\langle r,x|,
 \qquad F_r=\{x:(r,x)\in\Omega\}.
 \label{eq:fibre-definition}
\end{equation}
For an elementary kinematic matrix unit $E_{s,r}=|s\rangle\langle r|\otimes I_{R^c}$, a direct calculation gives the leakage operator
\begin{equation}
 (1-P)E_{s,r}P
   =\sum_{x\in F_r\setminus F_s}|s,x\rangle\langle r,x|.
 \label{eq:fibre-leakage}
\end{equation}
Equation~\eqref{eq:fibre-leakage} provides an exact characterization of physical preservation. Because commuting with $P$ requires the simultaneous preservation of $\mathcal H_0$ by an operator and its adjoint, applying the same calculation to the reverse matrix unit $E_{r,s}=E_{s,r}^\dagger$ yields the parallel criteria
\begin{equation}
 (1-P)E_{s,r}P=0\ \Longleftrightarrow\ F_r\subseteq F_s,
 \qquad
 [E_{s,r},P]=0\ \Longleftrightarrow\ F_r=F_s.
 \label{eq:fibre-criteria}
\end{equation}
Distinct triples $(s,r,x)$ label orthogonal matrix units on $\mathcal H$, so leakage from different coefficients in $\widehat A_R=\sum_{s,r}A_{s,r}E_{s,r}$ cannot cancel. Thus $\widehat A_R$ preserves $\mathcal H_0$ exactly when $F_r\subseteq F_s$ for each nonzero $A_{s,r}$. If $F_r=\varnothing$, then $E_{s,r}P=0$, so that unit has no physical action. On labels with nonempty fibres, inclusion is reflexive and transitive, giving the structural-matrix preorder.

The raw image $\mathcal A_0(R)$ is star-closed if and only if no strict inclusion occurs between nonempty fibres. To see necessity, suppose $\varnothing\neq F_r\subsetneq F_s$. The allowed unit $E_{s,r}$ induces a physical operator whose adjoint sends $|s,x\rangle$ to $|r,x\rangle$ for $x\in F_r$. Any supported lift of this adjoint must therefore have $(r,s)$ coefficient one. But for $y\in F_s\setminus F_r$, that same coefficient sends the physical state $|s,y\rangle$ out of $\mathcal H_0$ and must vanish. Hence no preserving lift exists.

Conversely, in the absence of strict inclusions, every allowed unit with nonzero physical action connects equal fibres. Its reverse unit is then a preserving lift of its physical adjoint, proving star closure. Reducing lifts connect equal fibres from the outset. Taking $C^*(\mathcal A_0(R))$ is a different operation: it adjoins physical adjoints whether or not they admit preserving lifts on $R$.

\subsection{An exact separation criterion}
\label{subsec:separation-criterion}

To separate a joint physical observable from the generated regional algebra $\mathcal A(R_1)\vee\mathcal A(R_2)$, we establish a single-matrix-element obstruction. Let $W$ be a Hermitian operator of unit operator norm supported on $R_1\cup R_2$ and commuting with the physical projector, $[W,P]=0$. Suppose there exist normalized physical states $\psi_0,\psi_1\in\mathcal H_0$ such that
\begin{equation}
 \langle\psi_1|W|\psi_0\rangle=1,\qquad
 \langle\psi_1|C|\psi_0\rangle=0
 \quad\text{for all }C\in\mathcal A(R_1)\vee\mathcal A(R_2).
 \label{eq:separating-transition}
\end{equation}
Under these conditions, the physical restriction $W_0=W\vert_{\mathcal H_0}$ satisfies the lower bound $\|W_0-C\|\geq|\langle\psi_1|(W_0-C)|\psi_0\rangle|=1$ for every $C\in\mathcal A(R_1)\vee\mathcal A(R_2)$. Because $\|W_0\|\leq\|W\|=1$ and $0\in\mathcal A(R_1)\vee\mathcal A(R_2)$, the lower bound is attained at $C=0$. Thus, in the operator norm,
\begin{equation}
 \operatorname{dist}\!\left(W_0,
             \mathcal A(R_1)\vee\mathcal A(R_2)\right)=1.
 \label{eq:distance-one}
\end{equation}
Since $W_0\in\mathcal A_0(R_1\cup R_2)$, Eq.~\eqref{eq:distance-one} proves strict inclusion in~\eqref{eq:additivity-inclusion}. Condition~\eqref{eq:separating-transition} rules out expressing $W_0$ as any linear combination of products of single-region physical operators and their adjoints on $\mathcal H_0$, not just as a single product. A physical measurement outcome is represented by an effect $F$, a positive contraction $0\le F\le I$ on $\mathcal H_0$; its probability in a given state is the expectation value of $F$.

\begin{corollary}[Relative-sign distinguishability]
\label{cor:relative-sign}
Let $W$ be a Hermitian operator of norm one supported on $R_1\cup R_2$ with $[W,P]=0$, and let $\psi_0,\psi_1\in\mathcal H_0$ be normalized physical states satisfying \eqref{eq:separating-transition}. Define the normalized superpositions
\[
 \psi_\pm=\frac{\psi_0\pm\psi_1}{\sqrt2}.
\]
These are mutually orthogonal physical states in $\mathcal H_0$. For every $C\in\mathcal A(R_1)\vee\mathcal A(R_2)$,
\[
 \langle\psi_+|C|\psi_+\rangle
 =\langle\psi_-|C|\psi_-\rangle,
 \qquad W_0\psi_\pm=\pm\psi_\pm.
\]
Consequently, any measurement whose effects belong to the generated regional algebra yields identical outcome statistics on $\psi_+$ and $\psi_-$, whereas the supported union observable $W$ distinguishes them deterministically.
\end{corollary}

\begin{proof}
Taking $C=I$ in~\eqref{eq:separating-transition} gives $\psi_0\perp\psi_1$. Star closure makes the reverse cross term vanish as well: $\langle\psi_0|C|\psi_1\rangle=\overline{\langle\psi_1|C^\dagger|\psi_0\rangle}=0$. Both superposition expectations are therefore the same half-sum of diagonal expectations. The restriction $W_0$ is a Hermitian contraction, and $\langle\psi_1|W_0|\psi_0\rangle=1$ saturates Cauchy--Schwarz, giving $W_0\psi_0=\psi_1$. Hermiticity gives the reverse unit matrix element, so the same argument yields $W_0\psi_1=\psi_0$. Thus $W_0\psi_\pm=\pm\psi_\pm$.
\end{proof}

The group constructions below use exterior fibres to resolve the permitted regional transitions; the Hopf construction tests the generated algebra directly.

\section{Local obstructions and the contact criterion}
\label{sec:local-criterion}

Disjoint additivity can fail without either region wrapping around the torus. In this section we fix a small square loop and vary only eight outer contact edges. The resulting criterion characterizes when the obstruction occurs and shows how the full union algebra is recovered at the same separation distance.

\subsection{Geometry and statement of the criterion}
We take a torus of size $N,M\ge8$ and let $L$ be the eight-edge boundary of the square with corners $(2,2),(4,2),(4,4),(2,4)$. At the four straight boundary vertices of $L$, the incident loop and exterior half-edges alternate cyclically, forming cusps; at the four corners, the two loop half-edges are cyclically adjacent. Loop configurations are written as counterclockwise tuples $b\in G^8$, with the physical group labels inverted along the top and left edges.

To define the surrounding support $S$, we take each exterior nearest-neighbor vertex of $V(L)$ and collect its three incident edges that do not meet $L$. This produces 28 edges vertex-disjoint from $L$. Let $\mathcal I$ be the set of four plaquettes inside $L$ and $\mathcal O$ the set of eight plaquettes outside $L$ that share an edge with $L$. The union of their boundaries, $D=\bigcup_{f\in\mathcal I\cup\mathcal O}\partial f$, consists of 32 edges and partitions $S$ into a 20-edge exterior base $S_0=S\setminus D$ and an 8-edge contact set $C=S\cap D$. We study the family of supports
\begin{equation}
 R_K=S_0\cup K,\qquad K\subseteq C.
 \label{eq:contact-family}
\end{equation}
Each inner face $f\in\mathcal I$ is paired with a two-edge subset $C_f\subset C$ lying across its adjacent outer faces. The four subsets $C_f$ partition $C$, with explicit coordinates given below. We say that $K$ covers $f$ if $K\cap C_f\ne\varnothing$. Two four-edge choices of $K$ are shown in Fig.~\ref{fig:local-criterion}. For every $K$, the regions $L$ and $R_K$ lie inside the coordinate square $[0,6]^2$, which is contained in a contractible disk on the torus, so neither support wraps a noncontractible cycle. Table~\ref{tab:result-scope} summarizes the geometric configurations and main results.

\begin{table}[!htbp]
\centering
\caption{Scope of the main results. Every statement uses the full magnetic physical space and original edge supports. The torus has $N$ vertices along each horizontal cycle and $M$ along each vertical cycle. The rows refer to the distinct geometries defined in the indicated sections.}
\label{tab:result-scope}
\small
\begin{tabular}{@{}p{.20\linewidth}p{.39\linewidth}p{.31\linewidth}@{}}
\hline
Input & Geometry & Result \\
\hline
Any finite group & Fixed eight-edge loop and eight-contact family, $N,M\ge8$ & Failure exactly for a non-Abelian group with all four inner faces covered; Section~\ref{sec:local-criterion}. \\
Any finite group & Same loop ($N,M\ge8$); arbitrary region avoiding its testing neighborhood & Full additivity; vertex distance at least two is sufficient; Section~\ref{sec:buffer-restoration}. \\
Any finite group & Fixed wrapping pair, $N,M\ge3$ & Complete block inclusion; equality exactly for Abelian groups; Section~\ref{sec:wrapping-classification}. \\
Kac--Paljutkin Hopf algebra & Specified vertex-disjoint supports on a $3\times3$ square torus & Norm-one obstruction to disjoint additivity; Section~\ref{sec:hopf-counterexample}. \\
\hline
\end{tabular}
\end{table}

Indexing each lattice plaquette by its lower-left vertex $(i,j)$, the exterior nearest-neighbor vertices of $V(L)$ that define $S$ are
$\{(i,1),(i,5):2\le i\le4\}\cup\{(1,j),(5,j):2\le j\le4\}$.
Each of these vertices is joined to $V(L)$ by a single edge, while its other three incident edges belong to $S$. In these coordinates, $\mathcal I$ consists of $(2,2),(3,2),(2,3),(3,3)$ and $\mathcal O$ consists of $(2,1),(3,1),(2,4),(3,4),(1,2),(1,3),(4,2),(4,3)$. Their boundary union $D$ has 32 edges, giving $|S|=28$, $|S_0|=20$, and the four contact pairs
\begin{align}
 C_{22}&=\{h_{2,1},v_{1,2}\},&C_{32}&=\{h_{3,1},v_{5,2}\},\nonumber\\
 C_{23}&=\{h_{2,5},v_{1,3}\},&C_{33}&=\{h_{3,5},v_{5,3}\}.
 \label{eq:explicit-pairs}
\end{align}
The configuration in Fig.~\ref{fig:local-criterion}(a) uses $K=\{h_{2,1},h_{3,1},h_{2,5},h_{3,5}\}$, whereas Fig.~\ref{fig:local-criterion}(b) uses $K=\{h_{2,1},v_{1,2},h_{3,1},v_{5,2}\}$. For an element $z\in Z(G)$ in the group center, the diagonal right action $b\mapsto bz$ multiplies each of the eight counterclockwise tuple entries by $z$. Because a physical edge label $g$ on the top or left boundary of $L$ is oriented so that it enters the counterclockwise cycle as $g^{-1}$, its transformed physical label is $(g^{-1}z)^{-1}=z^{-1}g=gz^{-1}$, where the final equality holds for every $g$ because $z$ is central.

\begin{theorem}[Local contact criterion]
\label{thm:local-criterion}
For every finite group $G$ and every subset $K\subseteq C$, the regional algebras $\mathcal A(R_K)$ and $\mathcal A(L)$ commute. Their join $\mathcal B_K=\mathcal A(R_K)\vee\mathcal A(L)$ satisfies
\begin{equation}
 \begin{gathered}
 \mathcal B_K\subsetneq\mathcal A(R_K\cup L)\\
 \Longleftrightarrow\quad
 \begin{cases}
 G\text{ is non-Abelian},\\
 K\cap C_f\ne\varnothing\quad\text{for all }f\in\mathcal I.
 \end{cases}
 \end{gathered}
 \label{eq:coverage-iff}
\end{equation}
When the condition on the right does not hold, $\mathcal B_K=\mathcal A(R_K\cup L)$. \end{theorem}

When $G$ is non-Abelian, the four independent conditions $K\cap C_f\ne\varnothing$ hold simultaneously for $3^4=81$ of the $2^8=256$ subsets $K\subseteq C$, and fail for the remaining 175 subsets. The smallest nonadditive supports in the family therefore have $|K|=4$, with one contact edge chosen from each $C_f$. Placing four contact edges on just two pairs $C_f$ restores equality. Thus neither edge count nor vertex distance suffices to determine whether the regional algebras generate the union algebra.

\subsection{Universal exterior tests and the loop algebra}
\label{subsec:local-loop-algebra}
To see how the obstruction arises, we first determine the physical transitions supported on the isolated loop $L$. Every loop configuration $b\in G^8$ extends to a global physical configuration on the lattice by choosing face potentials $p_f\in G$ and setting
\begin{equation}
 h_{i,j}=p_{i,j-1}^{-1}p_{i,j},\qquad
 v_{i,j}=p_{i,j}^{-1}p_{i-1,j}.
 \label{eq:local-potentials}
\end{equation}
With this parametrization, the magnetic word at every vertex cancels identically. Because each outer plaquette in $\mathcal O$ shares a single edge with $L$, choosing the four inner face potentials freely fixes the eight adjacent outer potentials through $p_o=p_f b_\epsilon^{-1}$. At each straight cusp of $L$, the inward-pointing exterior edge separates two inner plaquettes, so its label ranges over all of $G$ while the entire loop configuration $b$ is kept fixed. We use these potential configurations as explicit physical test backgrounds, rather than as a parametrization of all torus holonomy sectors.

Consider the bottom straight vertex $(3,2)$, where the counterclockwise tuple begins with $b_1=h_{2,2}$ and $b_2=h_{3,2}$. Writing $x=v_{3,2}$ for the interior edge and $y=v_{3,1}$ for the exterior edge, the vertex constraint is $b_2 x b_1^{-1}=y$. For a loop transition $|b'\rangle\langle b|$ to preserve $\mathcal H_0$ across every physical completion, it must satisfy
\begin{equation}
 b'_2x(b'_1)^{-1}=b_2xb_1^{-1}\quad\text{for all }x\in G.
 \label{eq:universal-center}
\end{equation}
Varying $x$ over $G$ forces the two incident loop labels to shift by the same central element, $(b'_2)^{-1}b_2=(b'_1)^{-1}b_1\in Z(G)$. While this local central constraint at a cusp was noted in Section~3 of Ref.~\cite{HolfesterSorce2026Gauss}, the potential completions above guarantee that $x$ actually ranges over $G$ at fixed $b$ for the closed square loop. As we verify next, the remaining vertex constraints propagate the same central shift around all eight edges of $L$, and any such uniform shift leaves every vertex word unchanged. The preserving matrix units on $L$ are therefore
\begin{equation}
 |bz\rangle\langle b|,\qquad b\in G^8,\quad z\in Z(G),
 \label{eq:local-loop-units}
\end{equation}
which follows once we trace the shift around the perimeter. At the straight cusp in Eq.~\eqref{eq:universal-center}, we have $b'_1=b_1z$ and $b'_2=b_2z$ with $z\in Z(G)$. At every other vertex of $L$, the magnetic constraint contains exactly two loop labels; with the exterior labels fixed, the shift on one loop edge uniquely determines that on its neighbor. To check the four corners explicitly, shift coordinates so that the lower-left corner of $L$ lies at $(0,0)$, and write the positively oriented loop edges as $A=h_{0,0}$, $B=h_{1,0}$, $C_{\rm v}=v_{2,0}$, $D'=v_{2,1}$, $E=h_{1,2}$, $F=h_{0,2}$, $J=v_{0,1}$, and $H=v_{0,0}$. The corner vertex constraints then read
\begin{align}
 AH&=v_{0,-1}h_{-1,0},&C_{\rm v}B^{-1}&=h_{2,0}^{-1}v_{2,-1},\nonumber\\
 D'E&=h_{2,2}v_{2,2},&F^{-1}J&=v_{0,2}h_{-1,2}^{-1}.
 \label{eq:corner-propagation}
\end{align}
Since the right-hand sides depend only on exterior edges, commuting the central element $z$ through each product shows that the counterclockwise tuple $(A,\allowbreak B,\allowbreak C_{\rm v},\allowbreak D',\allowbreak E^{-1},\allowbreak F^{-1},\allowbreak J^{-1},\allowbreak H^{-1})$ transforms uniformly as $b\mapsto bz$. A single unconstrained cusp therefore suffices to propagate the central shift around $L$, without requiring independent variations at all four cusps. Conversely, the straight-cusp and corner equations are all preserved by any uniform central shift $b\mapsto bz$. The reverse of $|bz\rangle\langle b|$ is $|b\rangle\langle bz|$, which is the unit in Eq.~\eqref{eq:local-loop-units} with input $bz$ and shift $z^{-1}$. Hence the raw algebra is already star-closed: $\mathcal A_0(L)=\mathcal A(L)$.

Each matrix unit in Eq.~\eqref{eq:local-loop-units} acts with the same scalar coefficient on every physical exterior completion, and their physical span forms the $C^*$-algebra $\mathcal A(L)=\mathcal A_0(L)$. The projections onto the orbits of $Z(G)$ in $G^8$ commute with every operator supported on the disjoint region $R_K$, and therefore commute with the generated join $\mathcal B_K$. Likewise, since kinematic lifts on disjoint supports commute on $\mathcal H_0$ and $\mathcal A_0(L)$ is star-closed, we have $[\mathcal A(R_K),\mathcal A(L)]=0$.

\subsection{A conditionally legal witness and separation from the join}
\label{subsec:local-witness}

Suppose now that $K$ covers all four inner faces and that $G$ is non-Abelian. Pick a noncentral element $t\notin Z(G)$ (for example, $t=(12)$ in $G=S_3$). Let $Q_K$ be the kinematic projector setting every edge in $R_K$ to the identity $e$, and let $T_L$ be the loop operator that swaps the uniform configurations $\bar e$ and $\bar t$ and annihilates all other configurations, where $\bar g=(g,\ldots,g)\in G^8$ denotes the counterclockwise tuple with constant label $g$. Their product on the kinematic space satisfies
\begin{equation}
 \begin{gathered}
 \widehat W_K=Q_K T_L,\\
 (1-P)\widehat W_KP=0,\qquad (1-P)T_LP\ne0.
 \end{gathered}
 \label{eq:conditional-product}
\end{equation}
To check that $\widehat W_K$ preserves the full magnetic subspace $\mathcal H_0$, note that any physical basis state restricts to a flat connection on the simply connected dual patch $0\le i,j\le5$. It therefore admits local face potentials on this patch regardless of the global holonomies around the torus. Fixing $p_{0,0}=e$, we set $p_{i,0}=p_{i-1,0}v_{i,0}^{-1}$ along the bottom row and define $p_{i,j}=p_{i,j-1}h_{i,j}$ vertically. Rewriting the vertex condition $C_{i,j}=e$ of Eq.~\eqref{eq:local-magnetic} as $v_{i,j}=h_{i,j}^{-1}v_{i,j-1}h_{i-1,j}$ then shows by induction on $j$ that Eq.~\eqref{eq:local-potentials} holds throughout the $6\times 6$ patch.

In this dual picture, the 20 edges of $S_0$ cross a closed cycle of 20 plaquettes:
$(0,1),\allowbreak (0,2),\allowbreak (0,3),\allowbreak (0,4),\allowbreak (1,4),\allowbreak (1,5),\allowbreak (2,5),\allowbreak (3,5),\allowbreak (4,5),\allowbreak (4,4),\allowbreak (5,4),\allowbreak (5,3),\allowbreak (5,2),\allowbreak (5,1),\allowbreak (4,1),\allowbreak (4,0),\allowbreak (3,0),\allowbreak (2,0),\allowbreak (1,0),\allowbreak (1,1)$.
Each contact edge in $C$ separates a plaquette on this 20-cycle from an outer plaquette in $\mathcal O$. On any physical configuration where $Q_K T_L$ does not vanish, the identity condition on $S_0$ equates all 20 cycle potentials to a common group element $p_*$. Because $K$ meets each pair $C_f$, at least one outer plaquette adjacent to each inner plaquette $f\in\mathcal I$ is also pinned to $p_*$. When the loop carries a uniform label $g\in\{e,t\}$, the loop edges joining those outer plaquettes to $f$ force all four inner potentials to equal $p_*g$, and the remaining loop edges then force all eight outer potentials in $\mathcal O$ to equal $p_*$. The four inward exterior edges at the cusps of $L$ consequently carry the identity label $e$. Swapping $g=e$ and $g=t$ alters only the eight edges of $L$ and leaves every vertex constraint satisfied. Since states outside the support of $Q_K T_L$ are mapped to zero, Eq.~\eqref{eq:conditional-product} holds on the entire physical space $\mathcal H_0$. 

We separate the physical restriction $W_K=\widehat W_K|_{\mathcal H_0}$ from the join $\mathcal B_K$ using the center-orbit projectors. Let $\psi_g\in\mathcal H_0$ be the physical basis state with uniform loop label $\bar g$ and identity labels on all edges outside $L$. Then
\begin{equation}
 \langle\psi_t|X|\psi_e\rangle=0\quad(X\in\mathcal B_K),
 \qquad \langle\psi_t|W_K|\psi_e\rangle=1.
 \label{eq:local-separation}
\end{equation}
Here $\langle\psi_t|X|\psi_e\rangle=0$ because $t\notin Z(G)$, so $\bar e$ and $\bar t$ lie in distinct $Z(G)$-orbits whose projectors commute with $\mathcal B_K$. Since $\|W_K\|=1$, the criterion of Section~\ref{subsec:separation-criterion} gives $\operatorname{dist}(W_K,\mathcal B_K)=1$, excluding $W_K$ from the norm closure of polynomials in operators from $\mathcal A(R_K)$ and $\mathcal A(L)$.

In measurement terms, let $\psi_\pm=(\psi_e\pm\psi_t)/\sqrt2$. By Corollary~\ref{cor:relative-sign}, every effect $0\le F\le I$ in $\mathcal B_K$ assigns the same expectation value to both superpositions, $\langle\psi_+|F|\psi_+\rangle=\langle\psi_-|F|\psi_-\rangle$, whereas $W_K$ has eigenvalues $\pm1$ on $\psi_\pm$. While superselection-like sector invisibility under restricted classes of observables is familiar in lattice gauge theories~\cite{VanAcoleyenEtAl2016Distillation}, here it witnesses a strict inclusion between algebras defined on the unmodified edge supports. Both $\mathcal A(R_K)$ and $\mathcal A(L)$ already contain all preserving operators on their supports without restricting to contractible generators, and the identical support pair is additive whenever $G$ is Abelian.

\subsection{An uncovered face restores the full algebra}
\label{subsec:uncovered-face}
When at least one inner face is left uncovered, the obstruction disappears and the full union algebra coincides with the join $\mathcal B_K$. To prove this, we use face relabelings on the kinematic space. Given face elements $k_f\in G$, the edge relabeling
\begin{align}
 h'_{i,j}&=k_{i,j-1}^{-1}h_{i,j}k_{i,j},\nonumber\\
 v'_{i,j}&=k_{i,j}^{-1}v_{i,j}k_{i-1,j}
 \label{eq:face-transport}
\end{align}
defines a kinematic permutation unitary $U(k)$ on $\mathcal H$. On unconstrained edge configurations, Eq.~\eqref{eq:face-transport} conjugates each vertex word according to
\begin{equation}
 C'_{i,j}=k_{i,j-1}^{-1}C_{i,j}k_{i,j-1}.
 \label{eq:transport-word}
\end{equation}
Hence $[U(k),P]=0$, so $U(k)$ maps $\mathcal H_0$ onto itself without fixing physical basis states pointwise. Suppose now that $f\in\mathcal I$ is uncovered, $K\cap C_f=\varnothing$. For any $a\in G$, set $k_f=a$, set $k_o=b_\epsilon a b_\epsilon^{-1}$ on the two outer faces adjacent to $f$, and set $k_{f'}=e$ on all other faces. Because $S_0$ avoids $D$ and the two contact edges modified by $k_o$ lie in $C_f\setminus K$, this unitary acts trivially on the joint configuration $(r,b)$ on $R_K\cup L$. At the same time, an inward exterior edge bordering $f$ is multiplied on one side by $a$ or $a^{-1}$, and therefore ranges over all of $G$ as $a$ varies. Any preserving matrix unit on $R_K\cup L$ with nonzero physical restriction must consequently pass the universal test of Eq.~\eqref{eq:universal-center} at the cusp bordering $f$. The corner-propagation relations in Eq.~\eqref{eq:corner-propagation} then force the entire loop transition to be a uniform central shift, which belongs to $\mathcal A(L)$.

To show that the remaining $R_K$ factor is also individually preserving, we use the boundary geometry of $R_K$.
\begin{lemma}[Cuspless factor]
\label{lem:cuspless-factor}
Suppose $R$ and $L$ are vertex-disjoint regions such that at every boundary vertex of $R$, the incident edges of $R$ form a single cyclically contiguous block. For any constraint-preserving joint matrix unit on $R\cup L$ with a nonzero restriction to $\mathcal H_0$, both its $R$ factor and the reverse $R$ matrix unit individually preserve $\mathcal H_0$.
\end{lemma}
\begin{proof}
Let $E_{r',r}E_{b',b}$ be a constraint-preserving joint matrix unit on $R\cup L$ satisfying $E_{r',r}E_{b',b}P\ne0$. For each vertex $v$ incident to $R$, the contiguity condition permits a cyclic permutation of the vertex constraint into the form $w_R(r)w_{\mathrm{ext}}(x)=e$, where $w_{\mathrm{ext}}(x)$ is the empty word whenever all four edges meeting $v$ belong to $R$. Because $R$ and $L$ are vertex-disjoint, $v$ is incident to no edge of $L$, so $E_{r',r}E_{b',b}$ leaves the exterior configuration $x$ at $v$ invariant. Since the input fibre is nonempty, preservation of $\mathcal H_0$ requires $w_R(r')=w_R(r)$. This equality is independent of $x$, ensuring that $E_{r',r}$ and its reverse $E_{r,r'}$ each preserve the vertex constraint at $v$ on every physical input. It remains to check that every region $R_K$ in the contact family satisfies the contiguity hypothesis. At any exterior nearest neighbor of $V(L)$, the three edges of $S$ form a contiguous block whose ends contain any incident contact edges; deleting unselected contacts therefore leaves a nonempty contiguous block. Every remaining vertex touched by $R_K$ is incident either to a single fixed edge of $S_0$ or to two cyclically adjacent fixed edges of $S_0$.
\end{proof}
When an inner face is uncovered, every preserving matrix unit on $R_K\cup L$ with nonzero physical restriction has a central loop factor in $\mathcal A(L)$, while Lemma~\ref{lem:cuspless-factor} shows that its $R_K$ factor and the reverse $R_K$ unit both preserve $\mathcal H_0$. This matrix-unit factorization captures the support-preserving part of the neutral-projection argument in Section~4.3 of Ref.~\cite{HolfesterSorce2026Gauss}. Because the scalar coefficient of each joint unit is independent of the exterior completion, taking linear combinations, products, and physical adjoints yields $\mathcal A(R_K\cup L)=\mathcal B_K$, completing the proof of Theorem~\ref{thm:local-criterion}. When $G$ is Abelian, a single physical completion already fixes the relative shift on $L$, which is automatically central, so the same factorization applies for every $K\subseteq C$.

\subsection{The factorwise-preservation inference}
\label{subsec:failed-inference}
The gap between joint and factorwise preservation can be stated directly in terms of exterior fibres. Writing $(r,b,x)$ for basis configurations on $R_K$, $L$, and the complement $(R_K\cup L)^c$, and $\Omega$ for the set of allowed magnetic configurations, a joint matrix unit preserves $\mathcal H_0$ when
\begin{equation}
 (r,b,x)\in\Omega\ \Longrightarrow\ (r',b',x)\in\Omega
 \qquad\hbox{for every }x.
 \label{eq:joint-conditional-test}
\end{equation}
By contrast, preservation of $\mathcal H_0$ by the loop factor $E_{b',b}$ alone requires
\begin{equation}
 (u,b,x)\in\Omega\ \Longrightarrow\ (u,b',x)\in\Omega
 \qquad\hbox{for every }u,x.
 \label{eq:loop-universal-test}
\end{equation}
Vertex disjointness of $R_K$ and $L$ does not imply that Eq.~\eqref{eq:joint-conditional-test} upgrades to Eq.~\eqref{eq:loop-universal-test}. In the proof of Proposition~4.3 in Ref.~\cite{HolfesterSorce2026Gauss}, orthogonality of distinct leakage terms validly reduces preservation of a sum to preservation of each joint matrix unit. However, the step immediately following Eq.~(4.60) of that work infers that each regional factor preserves $\mathcal H_0$ on its own. Comparing Eqs.~\eqref{eq:joint-conditional-test} and \eqref{eq:loop-universal-test} isolates the dropped quantifier over $u$: fixing $r$ on $R_K$ can restrict the admissible exterior configurations $x$ near $L$. When $K$ covers all four inner faces, the operator $\widehat W_K=Q_K T_L$ exploits this restriction to satisfy Eq.~\eqref{eq:joint-conditional-test} while $T_L$ violates Eq.~\eqref{eq:loop-universal-test}, and its unit distance from $\mathcal B_K$ rules out any alternative decomposition in the generated join. The theorems in Ref.~\cite{HolfesterSorce2026Gauss} for cuspless regions and collared supports do not rely on this step and remain valid.

\section{Restoration outside the testing neighborhood}
\label{sec:buffer-restoration}

Whereas the contact theorem of Section~\ref{sec:local-criterion} varies a specific family of peripheral supports, a second restoration theorem permits the accompanying region to have an arbitrary shape—including cusps—provided it avoids the neighborhood $D$ used for the loop tests. The additional step required for this general geometry is a transport of conditionally admissible transitions across loop inputs.

\begin{theorem}[Restoration outside the testing neighborhood]
\label{thm:local-buffer}
Let $G$ be any finite group, $N,M\ge8$, and let $L$ and $D$ be the fixed sets defined in Section~\ref{sec:local-criterion}. For any edge region $R$,

\begin{equation}
 R\cap D=\varnothing
 \quad\Longrightarrow\quad
 \mathcal A(R\cup L)=\mathcal A(R)\vee\mathcal A(L).
 \label{eq:local-buffer}
\end{equation}
\end{theorem}

Every vertex of $D$ lies at graph distance at most 1 from $V(L)$. Consequently, the hypothesis $R\cap D=\varnothing$ covers every edge region $R$ whose vertex set sits at graph distance at least 2 from $L$, for any finite group $G$. The outer 24-edge perimeter depicted in Fig.~\ref{fig:local-criterion}(c) furnishes an explicit realization. Note that Eq.~\eqref{eq:local-buffer} relies on the specific geometry of the square loop $L$ and does not assert additivity for arbitrary pairs of disjoint regions.

\begin{proof}
Because $R$ is disjoint from $D$, face relabelings supported entirely within $D$ generate all required exterior cusp variations without altering $R$. As in Section~\ref{sec:local-criterion}, these tests force the loop factor of any preserving union matrix unit with nonzero physical action to be an admissible central transition. Because $R$ may now contain cusps, however, Lemma~\ref{lem:cuspless-factor} does not apply, and a separate argument is needed to show that the $R$ factor is individually preserving. Let $E_R E_{b',b}$ be a preserving joint matrix unit with $E_R E_{b',b}P\ne0$. Multiplying by the admissible reverse loop unit $E_{b,b'}$ shows that the conditionally supported operator $E_R P_b$ preserves $\mathcal H_0$, where $P_b$ projects onto the normalized loop configuration $b$. Fix an arbitrary loop configuration $u\in G^8$. In Eq.~\eqref{eq:face-transport}, set the four inner face parameters to the identity ($k_f=e$) and set the eight outer face parameters to $k_{o(\epsilon)}=u_\epsilon b_\epsilon^{-1}$. Because each outer face shares exactly one edge with $L$, these eight assignments are mutually compatible. The resulting permutation unitary $U(k)$ is supported in $D$, commutes with $P$ and $E_R$, and maps the loop configuration $u$ to $b$ while leaving $R$ untouched. For this choice of $u$ and $b$, write $U=U(k)$ and let $P_u$ project onto loop configuration $u$. Conjugation gives the first identity, and summing over all loop configurations gives the second:
\begin{equation}
 U^\dagger(E_RP_b)U=E_RP_u,\qquad
 E_R=\sum_{u\in G^8}E_RP_u
 \label{eq:transport-factor}
\end{equation}
Since $U(k)$ preserves $\mathcal H_0$, the operator $E_R P_u$ preserves $\mathcal H_0$ for every $u\in G^8$, and summing over $u$ proves that the unrestricted operator $E_R$ preserves $\mathcal H_0$. Every preserving union matrix unit with nonzero physical action therefore factors into a product of individually preserving regional matrix units with an exterior-independent scalar coefficient. Taking linear spans and physical $C^*$-closures establishes $\mathcal A(R\cup L)=\mathcal A(R)\vee\mathcal A(L)$. In summary, additivity is restored by combining two mechanisms within $D$: varying the cusp edges to restrict the loop factor, and transporting loop configurations to uncondition the $R$ factor.
\end{proof}

Unlike group-averaging constructions~\cite{HarlowShaoSorceSrivastava2025Disjoint}, this proof uses $U(k)$ to relate conditional sectors within $\mathcal H_0$ rather than defining $\mathcal H_0$ as the $U(k)$-invariant subspace.

\subsection{Why outward enlargement does not remove the tests}
\label{subsec:outward-buffer-control}

The hypothesis of Theorem~\ref{thm:local-buffer} restricts the location of $R$ relative to the testing neighborhood $D$, rather than the total number of edges in $R$. To illustrate this distinction, one can enlarge the 24-edge outer perimeter of Fig.~\ref{fig:local-criterion}(c) by attaching its outward-pointing spokes. On a $10\times10$ torus, take the perimeter of the square $[0,6]^2$ and adjoin the horizontal edges $h_{9,j},h_{6,j}$ for $0\le j\le6$ together with the vertical edges $v_{i,9},v_{i,6}$ for $0\le i\le6$. The resulting 52-edge region $R_+$ remains disjoint from $D$ and maintains vertex-set distance 2 from $L$.

For a non-Abelian group $G$ and a noncentral element $t\notin Z(G)$, select $a\in G$ such that $at\ne ta$. In the potential formula of Eq.~\eqref{eq:local-potentials}, assign
\begin{equation}
 p_{2,2}=p_{2,1}=p_{1,2}=a,
 \qquad p_f=e\quad\hbox{on every other face}.
 \label{eq:buffer-counter-input}
\end{equation}
Because all vertex words telescope, this assignment defines a valid physical configuration in $\mathcal H_0$. Every edge of $L$ and every edge outside $D$ carries the identity label $e$, so the identity-label projector $Q_{R_+}$ on $R_+$ leaves this input state invariant. At the bottom cusp $(3,2)$, the incident loop labels are $b_1=b_2=e$, while the interior and exterior vertical edges carry $x=v_{3,2}=a$ and $y=v_{3,1}=a$. Applying the loop swap $T_L$ from the all-identity tuple $\bar e$ to the uniform tuple $\bar t$ changes the magnetic constraint word at $(3,2)$ to
\begin{equation}
 t a t^{-1}a^{-1}\ne e.
 \label{eq:buffer-commutator}
\end{equation}
Because $R_+$ is disjoint from $L$, the transition $T_L$ modifies no edge of $R_+$, and $Q_{R_+}$ also passes the resulting constraint-violating state. Consequently, the product $Q_{R_+}T_L$ fails to preserve $\mathcal H_0$. Moreover, because the input state defined by Eq.~\eqref{eq:buffer-counter-input} already carries the identity label on every edge outside $D$, projecting any additional edges outside $D$ onto $e$ leaves the state unchanged. Fixing exterior edges outside $D$ does not freeze the local degrees of freedom inside $D$ that obstruct noncentral loop transitions.

\section{Wrapping regions and the missing coherences}
\label{sec:group-counterexample}
\label{sec:wrapping-classification}

Noncontractible winding is not needed to violate disjoint additivity, as the local criterion already shows. Passing to a wrapping geometry, however, resolves the full algebra inclusion into finite matrix blocks and pinpoints the exact union observables that regional operators fail to generate.

\subsection{Geometry and regional algebras}
We work on an $N\times M$ torus ($N,M\ge3$) with the finite-group edge spaces, magnetic vertex constraints, and regional-algebra prescription of Section~\ref{sec:regional-algebras}. The physical subspace $\mathcal H_0$ enforces only the local magnetic vertex condition associated with the representation category $\operatorname{Rep}(G)$ of $G$~\cite{HolfesterSorce2026Gauss}: it is the full magnetic space rather than the fully projected topological ground space obtained by imposing both electric and magnetic projectors~\cite{BombinMartinDelgado2008NonAbelian}. Neither electric plaquette projections nor global-flux restrictions are imposed, and the analysis invokes no Hamiltonian or state-preparation protocol.

\begin{figure}[!htbp]
 \centering
 \includegraphics[width=\textwidth]{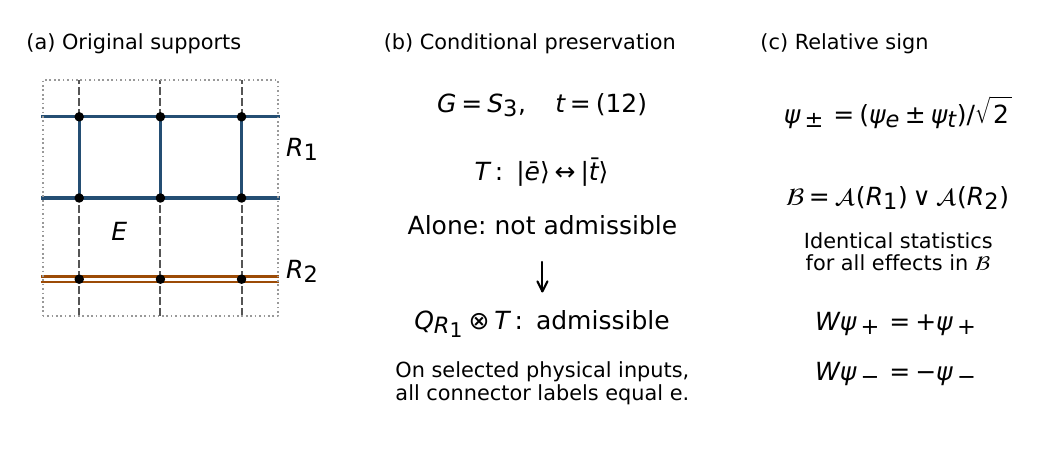}
 \caption{A joint observable distinguishes a relative sign even though the row exchange alone is not admissible. (a) On the periodic $3\times3$ torus, single solid lines mark $R_1$, double lines mark row $R_2$, and dashed lines mark the connecting edges $E$. Opposite dotted boundaries are identified. (b) For $G=S_3$, let $e$ be the identity and $t=(12)$. Write $|\bar g\rangle$ for the row basis state with every edge labeled $g$. The row operator $T$ swaps $|\bar e\rangle$ and $|\bar t\rangle$ and annihilates the other row basis states. The projector $Q_{R_1}$ selects the all-$e$ configuration on $R_1$. In the physical subspace it selects, the vertex constraints fix all connector labels to $e$ and require a constant row. Hence $W=Q_{R_1}\otimes T$ (identity on $E$ understood) preserves the physical subspace. (c) For $g=e,t$, the physical basis state $\psi_g$ has row $|\bar g\rangle$ and identity labels elsewhere [Eq.~\eqref{eq:group-separating-states}]. The states $\psi_\pm=(\psi_e\pm\psi_t)/\sqrt2$ give identical probabilities for every effect in $\mathcal B=\mathcal A(R_1)\vee\mathcal A(R_2)$, whereas $W\psi_\pm=\pm\psi_\pm$. The figure specifies a fixed observable, not an implementation protocol.}
 \label{fig:mechanism}
\end{figure}

The torus edges are divided into two vertex-disjoint regions $R_1$ and $R_2$ separated by an intermediate connector set $E$. Taking $R_2$ to be the horizontal cycle at row $j=0$, we assign to $R_1$ all remaining horizontal rows ($j\ne0$) together with the vertical edges linking rows $1$ through $M-1$, while $E$ comprises the vertical edges incident to row $0$:
\begin{align}
 R_2&=\{h_{i,0}\},\qquad
 E=\{v_{i,0},v_{i,M-1}\},\nonumber\\
 R_1&=\{h_{i,j}:j\ne0\}\cup\{v_{i,j}:1\leq j\leq M-2\}.
 \label{eq:region-partition}
\end{align}
The region $R_2$ is thus supported on a single unthickened noncontractible row [Fig.~\ref{fig:mechanism}(a)]. Although $R_1$ and $R_2$ share no vertices, $E$ bridges their boundaries: around every vertex of $R_2$, the two horizontal edges of $R_2$ alternate cyclically with the two vertical edges of $E$, forming a cusp on the square lattice.

For any edge subset $R$, the raw physical image $\mathcal A_0(R)$ and the regional $C^*$-algebra $\mathcal A(R)$ are given by Eqs.~\eqref{eq:raw-physical-image} and~\eqref{eq:regional-star-algebra}. The ensuing matrix-unit classification shows that $\mathcal A_0(R_1)$, $\mathcal A_0(R_2)$, and $\mathcal A_0(R_1\cup R_2)$ are already closed under physical adjoints for the partition~\eqref{eq:region-partition}; we do not assume star-closure for other regions. This decomposition determines both the regional join $\mathcal B=\mathcal A(R_1)\vee\mathcal A(R_2)$ and the union algebra $\mathcal A_U=\mathcal A(R_1\cup R_2)$ in full.

\subsection{Admissibility conditions for single-row operators}
The restriction on single-row operators is already visible for $G=S_3$: because $Z(S_3)$ is trivial, every operator in $\mathcal A(R_2)$ is diagonal in the group basis, even though the union algebra contains off-diagonal observables. For an arbitrary finite group $G$, fix a row configuration $b_i=h_{i,0}$. Any assignment of adjacent vertical labels $c_i=v_{i,0}$ extends to a physical configuration on the torus via
\begin{align}
 h_{i,0}&=b_i,\quad h_{i,1}=b_i^{-1},\quad h_{i,j}=e\ (j\geq2),\nonumber\\
 v_{i,0}&=c_i,\quad v_{i,j}=d_i\ (j\geq1),\quad
 d_i=b_i c_i b_{i-1}^{-1}.
 \label{eq:arbitrary-row-extension}
\end{align}
This assignment satisfies the vertex constraint at every site, including at the minimal height $M=3$, so the exterior labels $c_i$ may be varied freely across $G^N$.

If an admissible lift on $R_2$ has a nonzero matrix element $A_{b',b}$, it must preserve every physical completion of $b$. Because distinct output rows $b'$ are orthogonal, constraint violations from one transition $(b,b')$ cannot be cancelled by other terms, so $(b,b')$ must satisfy
\begin{equation}
 b'_i c_i(b'_{i-1})^{-1}=b_i c_i b_{i-1}^{-1}
 \quad(\forall i\in\mathbb Z_N,\ \forall c_i\in G).
 \label{eq:row-environment-test}
\end{equation}
Taking $c_i=e$ gives $b'_i=b_i z$ for an $i$-independent element $z\in G$, and varying $c_i\in G$ then requires $z c_i z^{-1}=c_i$, so $z\in Z(G)$. Conversely, multiplying every entry of $b$ by a common central element $z$ preserves all vertex constraints for every exterior configuration. Hence
\begin{equation}
 |b'\rangle\langle b|\otimes I\in K_{R_2}
 \quad\Longleftrightarrow\quad b'=bz,\qquad z\in Z(G).
 \label{eq:row-central-shift}
\end{equation}
Section~3 of Ref.~\cite{HolfesterSorce2026Gauss} derived this uniform central-shift condition assuming exterior extendibility; Eq.~\eqref{eq:arbitrary-row-extension} supplies the required completions for every row $b$ and adjacent tuple $c$ on the torus. Because every $b\in G^N$ has a nonempty exterior fiber, Eq.~\eqref{eq:row-central-shift} determines $\mathcal A_0(R_2)$ completely and shows that it is star-closed, decomposing into blocks on the diagonal center orbits in $G^N$. When $Z(G)$ is trivial, as for $S_3$, these orbits are singletons and every admissible operator on $R_2$ is diagonal.

\subsection{Mechanism for joint admissible observables}
Choose a noncentral element $t\notin Z(G)$ (for example, $t=(12)$ in $S_3$), write $\bar g=(g,\ldots,g)$ for a constant row configuration, and set
\begin{align}
 Q_{R_1}&=|\mathbf e_{R_1}\rangle\langle\mathbf e_{R_1}|,\nonumber\\
 T_{R_2}&=|\bar t\rangle\langle\bar e|+|\bar e\rangle\langle\bar t|,
 \qquad W=Q_{R_1}\otimes T_{R_2}\otimes I_E.
 \label{eq:group-joint-witness}
\end{align}
On $R_2$ alone, $T_{R_2}$ is not admissible: in $S_3$, taking $c_i=(123)$ violates Eq.~\eqref{eq:row-environment-test} because $t c_i t^{-1}\ne c_i$.

In the joint operator $W$, the projector $Q_{R_1}$ restricts the physical states on which $T_{R_2}$ acts. Namely, $Q_{R_1}$ commutes with $P$ and fixes every edge of $R_1$ to $e$ on $\operatorname{ran}(Q_{R_1}P)$. The vertex constraints along rows $1$ and $M-1$ then force every connector edge in $E$ to equal $e$, reducing the constraints along row $0$ to $b_i=b_{i-1}$, which holds for every constant row $\bar g$. Consequently, $W$ exchanges the physical states with row configurations $\bar e$ and $\bar t$, annihilates all other constant rows, and vanishes where $Q_{R_1}=0$. Thus $W$ preserves $\mathcal H_0$, and since $W$ is Hermitian with $\|W\|=1$, it satisfies $[W,P]=0$.

Even a single matrix-unit term of $W$ refutes the inference from joint admissibility to factorwise admissibility. With $\widehat A_1=Q_{R_1}\otimes I$ and $\widehat A_2=I_{R_1}\otimes|\bar t\rangle\langle\bar e|\otimes I_E$,
\begin{equation}
 \begin{gathered}
 (1-P)\widehat A_1\widehat A_2P=0\\
 \nRightarrow\quad (1-P)\widehat A_iP=0\quad\text{for each }i=1,2.
 \end{gathered}
 \label{eq:failed-factor-inference}
\end{equation}
This factorwise step is invoked in the proof of Proposition~4.3 of Ref.~\cite{HolfesterSorce2026Gauss} immediately after Eq.~(4.60). Whereas $\widehat A_2\in K_{R_2}$ requires $\widehat A_2$ to preserve every physical completion of $\bar e$, the factor $\widehat A_1$ projects out the completions on which the transition $\bar e\to\bar t$ violates the Gauss law; $\widehat A_1$ is admissible, $\widehat A_2$ is not, and $Q_{R_1}$ enters as part of the observable $W$ rather than as a measurement postselection. While Eq.~\eqref{eq:failed-factor-inference} rules out factorwise admissibility for this decomposition, showing that $W|_{\mathcal H_0}\notin\mathcal B$ under any generated presentation requires an invariant of the full algebra $\mathcal B$.

\subsection{Strict separation from the generated algebra}
Let $J_{[b]}$ be the orthogonal projector on $\mathcal H_0$ onto configurations whose $R_2$ tuple lies in the diagonal center orbit $[b]=bZ(G)$. By Eq.~\eqref{eq:row-central-shift}, every operator in $\mathcal A_0(R_2)$ preserves $[b]$ and therefore commutes with $J_{[b]}$. Every operator in $\mathcal A_0(R_1)$ acts trivially on $R_2$ and likewise commutes with $J_{[b]}$. Because $J_{[b]}$ is self-adjoint on $\mathcal H_0$, it also commutes with their physical adjoints, and hence with the entire generated star algebra $\mathcal B$.

Consider the normalized physical basis states
\begin{equation}
 \psi_g=|\mathbf e_{R_1}\rangle\otimes|\bar g\rangle
                   \otimes|\mathbf e_E\rangle,\qquad g=e,t.
 \label{eq:group-separating-states}
\end{equation}
The condition $t\notin Z(G)$ places $\psi_e$ and $\psi_t$ in distinct center orbits. Commutation with $J_{[b]}$ then gives
\begin{equation}
 \langle\psi_t|C|\psi_e\rangle=0\quad(C\in\mathcal B),
 \qquad \langle\psi_t|W|\psi_e\rangle=1.
 \label{eq:group-separation}
\end{equation}
By the criterion of Section~\ref{subsec:separation-criterion}, $\|W|_{\mathcal H_0}-C\|\geq1$ for all $C\in\mathcal B$, and $C=0$ saturates the bound to give $\operatorname{dist}(W|_{\mathcal H_0},\mathcal B)=1$.

\subsection{Complete regional and union algebras}
The invariant $J_{[b]}$ is part of a complete block decomposition of $\mathcal B\subseteq\mathcal A_U$. We record the group labels on the two sets of vertical connecting edges by the pair of boundary tuples $\xi=(c,d)\in G^N\times G^N$, where $c_i=v_{i,0}$ and $d_i=v_{i,M-1}$. For these sector data, we define
\begin{align}
 Y_\xi&=\{b\in G^N:b_i c_i b_{i-1}^{-1}=d_i\ \text{for all }i\},\nonumber\\
 C_\xi&=c_{N-1}\cdots c_0,\qquad
 \kappa_\xi=|C_G(C_\xi)|,\qquad z_G=|Z(G)|.
 \label{eq:sector-data}
\end{align}
Here $C_G(C_\xi)$ is the centralizer of the cycle holonomy $C_\xi$ in $G$. Let $X_\xi$ be the set of $R_1$ configurations compatible with $\xi$ and all interior vertex constraints of $R_1$, and let $\Xi$ be the set of active sectors $\xi$ for which both $X_\xi$ and $Y_\xi$ are nonempty. For every $\xi\in\Xi$, $|Y_\xi|=\kappa_\xi$ and $m_\xi:=|X_\xi|=|G|^{N(M-2)}\kappa_\xi$. These active sectors partition the full magnetic subspace $\mathcal H_0$ without any external superselection rule, and the free right action of the diagonal center $Z(G)$ partitions each $Y_\xi$ into $\kappa_\xi/z_G$ orbits of size $z_G$.

\begin{theorem}[Represented wrapping algebras]
\label{thm:wrapping-blocks}
For the fixed regions in Eq.~\eqref{eq:region-partition}, $N,M\ge3$, and any finite group $G$, under the physical basis identification $|r,b,\xi\rangle$, the represented algebras decompose as
\begin{align}
 \mathcal H_0&\cong\bigoplus_{\xi\in\Xi}
             \mathbb C^{m_\xi}\otimes\mathbb C^{\kappa_\xi},\nonumber\\
 \mathcal B&\cong\bigoplus_{\xi\in\Xi}
             \bigoplus_{O\in Y_\xi/Z(G)}M_{m_\xi z_G}(\mathbb C),\nonumber\\
 \mathcal A_U&\cong\bigoplus_{\xi\in\Xi}M_{m_\xi\kappa_\xi}(\mathbb C).
 \label{eq:complete-blocks}
\end{align}
\end{theorem}

Within each sector block $M_{m_\xi\kappa_\xi}(\mathbb C)$ of $\mathcal A_U$, $\mathcal B$ is the block-diagonal subalgebra preserving the center orbits $O\in Y_\xi/Z(G)$. In sector $\xi$, $\mathcal A(R_1)$ and $\mathcal A(R_2)$ act on the respective tensor factors $\mathbb C[X_\xi]$ and $\mathbb C[Y_\xi]$, so they and their physical adjoints mutually commute. Although an operator in $\mathcal A(R_2)$ acts with identical matrix elements on a given orbit $O$ across every sector $\xi$ with $O\subseteq Y_\xi$, the projector onto each sector $\xi$ is an admissible operator supported on $R_1$ alone; multiplying by these sector projectors separates the sectors and generates all block matrices in Eq.~\eqref{eq:complete-blocks}. The proof of this reverse inclusion, the sector counts, and the star closure of the raw images are given in Appendix~\ref{app:wrapping-details}.

When $G$ is Abelian, $\kappa_\xi=z_G=|G|$, so each $Y_\xi$ consists of a single center orbit and $\mathcal B=\mathcal A_U$. When $G$ is non-Abelian, every $g\in G$ satisfies $|C_G(g)|>|Z(G)|$: indeed, $|C_G(g)|=|Z(G)|$ would force $C_G(g)=Z(G)$ since $Z(G)\subseteq C_G(g)$, which implies $g\in Z(G)$ and hence $C_G(g)=G$, contradicting that $G$ is non-Abelian. Therefore,
\begin{equation}
 \begin{gathered}
 \mathcal B=\mathcal A_U
 \quad\Longleftrightarrow\quad G\ \text{is Abelian},\\
 \text{for the regions in Eq.~\eqref{eq:region-partition}}.
 \end{gathered}
 \label{eq:additivity-iff}
\end{equation}
For non-Abelian $G$, the operators missing from $\mathcal B$ are precisely the inter-orbit coherences within each active connector sector $\xi$; all intra-orbit transitions and orbit populations lie in $\mathcal B$, and neither $\mathcal B$ nor $\mathcal A_U$ couples distinct sectors $\xi\ne\xi'$. Equation~\eqref{eq:additivity-iff} is a classification for the fixed partition~\eqref{eq:region-partition}, not a statement about arbitrary region geometries.

To isolate the witness $W$ inside Theorem~\ref{thm:wrapping-blocks}, define $\psi_g$ by Eq.~\eqref{eq:group-separating-states} for every $g\in G$ and let $Q=(Q_{R_1}\otimes I)|_{\mathcal H_0}$. Then $Q\mathcal H_0=\operatorname{span}\{\psi_g:g\in G\}$ is contained in the trivial connector sector $\xi=(e,e)$, in which both components are uniform identity tuples in $G^N$. Compressing by $Q$ yields
\begin{equation}
 Q\mathcal BQ\cong\bigoplus_{gZ(G)\in G/Z(G)}M_{z_G}(\mathbb C),
 \qquad Q\mathcal A_UQ\cong M_{|G|}(\mathbb C).
 \label{eq:witness-corners}
\end{equation}
Since $Q\in\mathcal B$, both $Q\mathcal BQ$ and $Q\mathcal A_UQ$ are unital corner subalgebras with identity $Q$. For any nontrivial group $G$, $Q\mathcal H_0$ does not reduce $\mathcal B$ and the compression $X\mapsto QXQ$ is not a star homomorphism on $\mathcal B$ (an explicit partial isometry in Appendix~\ref{app:wrapping-details} witnesses the failure of multiplicativity). Nevertheless, for any physical state $\rho$ supported on $Q\mathcal H_0$, $\operatorname{Tr}(\rho X)=\operatorname{Tr}(\rho QXQ)$ so the corners~\eqref{eq:witness-corners} determine all measurement statistics on $Q\mathcal H_0$. For $G=S_3$, $Q\mathcal BQ\cong\mathbb C^6$ consists of operators diagonal in $\{\psi_g:g\in S_3\}$, whereas $Q\mathcal A_UQ\cong M_6(\mathbb C)$, and $W|_{\mathcal H_0}$ is the Hermitian swap between $\psi_e$ and $\psi_t$, namely the sum of the two opposite off-diagonal matrix units.

Throughout this comparison, $\mathcal A(R_1)$ and $\mathcal A(R_2)$ include the restrictions of all admissible lifts supported on the full regions $R_1$ and $R_2$ rather than only on contractible patches, with $R_2$ itself wrapping the torus. All diagonal operators on the row configuration belong to $\mathcal A(R_2)$; what $\mathcal B$ lacks for non-Abelian $G$ are the transitions between distinct center orbits at fixed connector labels $\xi$, without any external projection onto global flux sectors. Because this same wrapping partition obeys $\mathcal B=\mathcal A_U$ for every Abelian group $G$, the failure of disjoint additivity cannot be attributed to noncontractible support alone.

Appendix~\ref{sec:wrapping-variants} gives two further controls for the wrapping geometry, with supports and lattice sizes specified there. The construction with direct adjacency along only one boundary of $R_2$ still violates additivity for every finite non-Abelian group; the specified strip with exterior buffer rows on both sides is additive for all finite groups.

\section[A regular-lattice counterexample for the Kac--Paljutkin algebra]{A regular-lattice counterexample for \texorpdfstring{$H_8$}{H8}}
\label{sec:hopf-counterexample}

In the group model, magnetic constraints are diagonal in a configuration basis, so preserving extensions can be tested entry by entry. For the finite-dimensional Hopf $C^*$-algebra $H_8$, the vertex projectors are no longer diagonal in a group basis and are instead expressed as invariant tensors in a Peter--Weyl basis~\cite{BuerschaperEtAl2013Hierarchy}. Quantum-double and string-net models~\cite{Kitaev2003FaultTolerantAnyons,LevinWen2005StringNet} are linked through local Hilbert-space completions~\cite{BuerschaperAguado2009Mapping}. Electric--magnetic dualities trade the Hopf algebra and lattice for their duals~\cite{BuerschaperEtAl2013EMDuality}, while further correspondences identify protected and excited sectors with Turaev--Viro state-sum data~\cite{BalsamKirillov2012KitaevTV}. At the boundary, Kitaev and Kong construct gapped string-net models from indecomposable module categories, with boundary excitations represented by modules over finite-dimensional $C^*$ weak-Hopf algebras~\cite{KitaevKong2012GappedBoundaries}. Even so, bulk topological order does not by itself fix the boundary net: string-net models sharing the same bulk phase can support non-isomorphic boundary quasi-local algebras~\cite{JonesEtAl2025BoundaryAlgebras}.

Certain lattice correspondences also track regional operator algebras directly. In Levin--Wen models constructed by gauging tube-algebra symmetry~\cite{KawagoeEtAl2024LevinWenGauge}, conjugation by the gauging unitary maps the chosen symmetric algebra on each bounded contractible patch onto the commutant of the boundary tube action on the constrained patch space. For a rectangular patch, this commutant is contained in the boundary-label-preserving algebra, which in turn embeds in the commutant on the rectangle enlarged by one layer. Exact unitary conjugation of the chosen net goes beyond a global state-space map, while the bounded-spread intertwining shifts microscopic supports rather than identifying algebras on the same edge set. Addressing disjoint additivity in the magnetic model additionally requires a regional dictionary for the full finite-torus magnetic subspace $\mathcal H_0$: for each fixed subset of original lattice edges, one must identify the physical restrictions of all constraint-preserving lifts and the $C^*$-algebra they generate, with adjoints taken on $\mathcal H_0$.

A complementary algebraic framework is provided by Hopf lattice gauge theory. On ciliated ribbon graphs, Meusburger and Wise construct gauge theory for finite-dimensional ribbon Hopf algebras~\cite{MeusburgerWiseHopfGauge}, and for semisimple Hopf input Meusburger links the Kitaev model to combinatorial quantization of the Drinfeld double, matching the gauge-invariant observable algebras and, once flatness is imposed, identifying the operator algebra on the protected space with the quantum moduli algebra~\cite{Meusburger2017KitaevGauge}. Below we build the needed fixed-support regional reduction directly on the regular square torus while keeping all unprojected electric sectors inside $\mathcal H_0$. In this calculation, edge orientations and endpoint orderings play the same structural role that they do for Hopf ribbon operators~\cite{YanChenCui2022HopfRibbons}.

Our proof on the regular $3\times3$ torus constructs a supported joint observable for $H_8$ on $R_1\cup R_2$ and separates it from the full star algebra $\mathcal A(R_1)\vee\mathcal A(R_2)$. The main technical step identifies a reducing subspace of the magnetic physical space $\mathcal H_0$ with the entire magnetic space of an auxiliary two-vertex ribbon graph, which parametrizes physical vectors inside the torus lattice while keeping the ambient edge set and regional supports fixed. The auxiliary projection is supported outside $R_2$ and commutes with $P$, so every constraint-preserving lift on $R_2$, together with its physical adjoint, reduces the selected subspace. Surjectivity of the graph parametrization transports both products and physical adjoints alongside individual generators. Within this compressed picture, preserving row lifts carry vanishing matrix blocks between the four one-dimensional representation labels and the two-dimensional label. Combined with an endpoint-sector factorization showing that the two regional star algebras commute on $\mathcal H_0$, this block structure rules out approximating the joint witness within the generated algebra.

\subsection{The model and the joint observable}
\label{subsec:hopf-model-witness}

Take $K=H_8$ as the vertex Hopf algebra and $H=K^\vee$ as the edge algebra, with magnetic vertex projectors defined from the normalized Haar integral of $K$ \cite{BuerschaperEtAl2013Hierarchy,HolfesterSorce2026Gauss}. The irreducible representations of $K$ are indexed by five labels with dimensions
\begin{equation}
 r\in\{0,\eta,s,\bar s,V\},\qquad
 (d_0,d_\eta,d_s,d_{\bar s},d_V)=(1,1,1,1,2).
 \label{eq:hopf-labels}
\end{equation}
Here $0$ denotes the trivial representation. For each label $r$, let $\rho_r:K\to\operatorname{End}(V_r)$ be the representation on its $d_r$-dimensional carrier. Its matrix coefficients are the linear functionals $Z^r_{ij}\in K^\vee$ given by $Z^r_{ij}(a)=\langle i|\rho_r(a)|j\rangle$ for $a\in K$. The indices run over an orthonormal basis of $V_r$; we use $0,1$ for the two-dimensional representation. Each edge carries the orthonormal endpoint-ordered Peter--Weyl basis
\begin{equation}
 |r,i,j\rangle_{\rm end}=\sqrt{d_r}\,Z^r_{ji},
 \label{eq:hopf-endpoint-basis}
\end{equation}
in which the outgoing endpoint index $i$ precedes the incoming index $j$, reversing the internal index order of the matrix coefficients $Z^r_{ij}$ used in \cite{HolfesterSorce2026Gauss}. We drop the subscript ${\rm end}$ from now on; explicit formulas for the Hopf structure maps, Haar inner product, and vertex projectors are collected in Appendix~\ref{app:hopf-data}.

On the $3\times3$ torus equipped with the partition \eqref{eq:region-partition}, we direct horizontal edges eastward in rows one and two and westward in row zero. Vertical edges point northward apart from $v_{0,0}$, which is directed southward from $u=(0,1)$ to $v=(0,0)$. Around every vertex, incident half-edges follow the counterclockwise compass order east, north, west, south. The cilium marking the start of the local cyclic order sits at the outgoing horizontal edge for all vertices in rows zero and one, at the northward edge for $(0,2)$, and at the eastward edge for $(1,2)$ and $(2,2)$. Throughout, only the magnetic vertex projectors $P_v$ are imposed.

\begin{theorem}
\label{thm:hopf-obstruction}
For the $H_8$ magnetic model on the regular $3\times3$ square torus, disjoint additivity fails for the vertex-disjoint regions $R_1$ and $R_2$: there exists a norm-one Hermitian operator $W_{H_8}$ supported on $R_1\cup R_2$ satisfying $[W_{H_8},P]=0$ whose physical restriction $W_{H_8}\vert_{\mathcal H_0}$ has distance one from $\mathcal A(R_1)\vee\mathcal A(R_2)$. This obstruction holds without assuming that the raw regional physical images are star-closed.
\end{theorem}

Writing $|0\rangle$ for the normalized basis state of the trivial label $r=0$ on one edge, we introduce the trivial-label projector on $R_1$,
\begin{equation}
 Q_{R_1}=|\mathbf0_{R_1}\rangle\langle\mathbf0_{R_1}|.
 \label{eq:hopf-R1-projection}
\end{equation}
Identity tensor factors on complementary edges are left implicit when regional projectors act on $\mathcal H$. Every vertex projector $P_v$ preserves the Peter--Weyl representation labels on its incident edges, giving $[Q_{R_1},P]=0$. For any state in $Q_{R_1}\mathcal H_0$, each boundary vertex of $R_1$ has three incident edges in $R_1$ fixed to the trivial representation, so a nonzero vertex invariant exists only when the fourth incident edge—lying in $E$—likewise carries label $0$. Every edge of $E$ is therefore fixed to $0$ on $Q_{R_1}\mathcal H_0$.

Once all edges of $E$ carry label $0$, the three vertex constraints along row zero reduce to bivalent singlet projections. Schur's lemma (Appendix~\ref{subsec:hopf-path-isometry}) then forces all three edges of $R_2$ to share a single representation label $r$, with one normalized closed-loop vector for each label. Labeling the westward row-zero cycle in sequence as $\epsilon_1=h_{2,0}$, $\epsilon_2=h_{1,0}$, $\epsilon_3=h_{0,0}$, that loop state reads
\begin{equation}
 |\omega_r\rangle
 =d_r^{-3/2}\sum_{i_1,i_2,i_3}
 |r,i_1,i_2\rangle_{\epsilon_1}\otimes
 |r,i_2,i_3\rangle_{\epsilon_2}\otimes
 |r,i_3,i_1\rangle_{\epsilon_3}.
 \label{eq:hopf-loop-state}
\end{equation}
Five orthonormal vectors consequently span the projected physical subspace:
\begin{equation}
 Q_{R_1}\mathcal H_0
 =\operatorname{span}\{\Psi_r:r=0,\eta,s,\bar s,V\},
 \qquad
 \Psi_r=|\mathbf0_{R_1}\rangle\otimes|\omega_r\rangle
                         \otimes|\mathbf0_E\rangle.
 \label{eq:hopf-vacuum-sector}
\end{equation}
Our construction pairs the trivial label $0$ with the two-dimensional label $V$. Testing row operators against nontrivial exterior states in Section~\ref{subsec:hopf-supported-compression} shows that every constraint-preserving lift on $R_2$ has zero physical matrix element between $\Psi_0$ and $\Psi_V$. Within $Q_{R_1}\mathcal H_0$, however, all exterior edges carry label $0$, so swapping $|\omega_0\rangle$ and $|\omega_V\rangle$ respects every vertex constraint, and multiplying that swap by $Q_{R_1}$ confines its action on $\mathcal H_0$ to $Q_{R_1}\mathcal H_0$.

We therefore set the joint witness on $R_1\cup R_2$ to be
\begin{equation}
 W_{H_8}=Q_{R_1}\otimes
 \bigl(|\omega_V\rangle\langle\omega_0|
       +|\omega_0\rangle\langle\omega_V|\bigr)\otimes I_E.
 \label{eq:hopf-joint-witness}
\end{equation}
Within the reducing subspace $Q_{R_1}\mathcal H_0$, the operator $W_{H_8}$ swaps $\Psi_0$ with $\Psi_V$ and sends $\Psi_\eta$, $\Psi_s$, and $\Psi_{\bar s}$ to zero, while on the orthogonal physical complement where $Q_{R_1}=0$ it vanishes identically. It follows that $W_{H_8}$ leaves $\mathcal H_0$ invariant, has operator norm one, and by Hermiticity satisfies $[W_{H_8},P]=0$, with transition amplitude $\langle\Psi_V|W_{H_8}|\Psi_0\rangle=1$.

\subsection{A reducing subspace for every row lift}
\label{subsec:hopf-supported-compression}

\begin{figure}[!htbp]
 \centering
 \includegraphics[width=\textwidth]{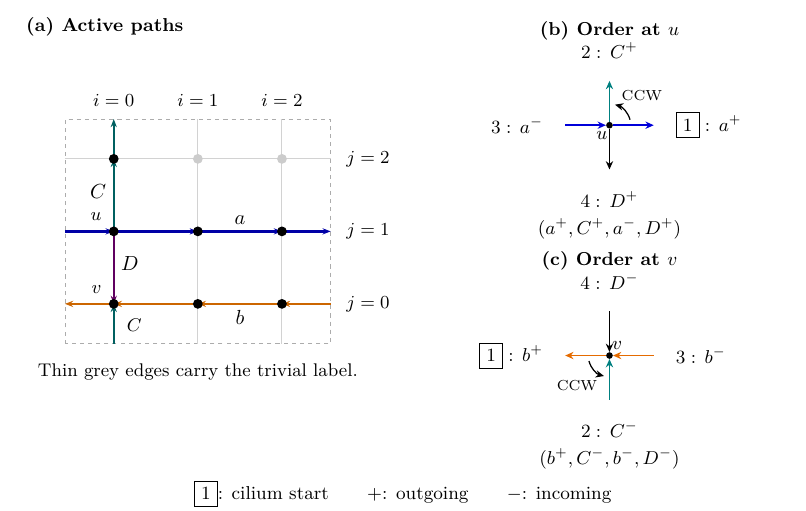}
 \caption{(a) Active subpaths for the $H_8$ subspace reduction on the periodic $3\times3$ square torus, with opposite dotted boundaries identified. The horizontal loops $a$ and $b$ are directed east and west, whereas $C$ and $D$ both travel from $u$ to $v$, heading north across the periodic boundary and south along a single edge, respectively. Applying $Q_{\rm aux}$ projects every thin grey edge onto the trivial label $0$, leaving four active incident half-edges at $u$ and at $v$ while rendering all other active vertices bivalent. (b,c) Local cyclic order of the active half-edges at $u$ and $v$. The boxed $1$ marks the cilium start, and the numerals $1$--$4$ together with the curved arrow trace the counterclockwise ordering in Eq.~\eqref{eq:hopf-active-orders}. Superscripts $+$ and $-$ denote outgoing and incoming half-edges relative to the displayed vertex; all remaining vertex cilia follow Section~\ref{subsec:hopf-model-witness}.}
 \label{fig:hopf-reduction}
\end{figure}

To probe row lifts in nontrivial exterior backgrounds, we retain the three-edge westward loop $b=R_2$ along row zero and link it to the three-edge eastward loop $a$ along row one as illustrated in Figure~\ref{fig:hopf-reduction}(a). Two directed paths run from $u$ to $v$: a two-edge path $C$ heading north along $v_{0,1}$ and $v_{0,2}$ across the periodic boundary, and a single-edge path $D$ heading south along $v_{0,0}$. Let $Q_{\rm aux}$ project every edge outside $a\cup b\cup C\cup D$ onto the trivial representation $0$. Unlike $Q_{R_1}\mathcal H_0$, the subspace $Q_{\rm aux}\mathcal H_0$ admits nontrivial representation labels along $a$, $C$, and $D$. Since $b=R_2$, the support of $Q_{\rm aux}$ lies inside $R_2^c$, and label preservation at every vertex ensures $[Q_{\rm aux},P]=0$.

Dropping the trivial tensor factors from the vertex projectors leaves four active incident half-edges at $u$ and at $v$, with local cyclic orderings
\begin{equation}
 u:(a^+,C^+,a^-,D^+),\qquad
 v:(b^+,C^-,b^-,D^-).
 \label{eq:hopf-active-orders}
\end{equation}
Every remaining active vertex becomes bivalent, where the magnetic projector enforces equality of the two adjacent edge labels and contracts their shared endpoint index. Along a directed $m$-edge path, the vectors satisfying all internal bivalent constraints form the range of the path isometry
\begin{equation}
 J_m|r,i,j\rangle
 =d_r^{-(m-1)/2}
 \sum_{k_1,\ldots,k_{m-1}}
 |r,i,k_1\rangle\otimes\cdots\otimes|r,k_{m-1},j\rangle.
 \label{eq:hopf-path-map}
\end{equation}
When $m=1$, $J_1$ is the identity operator. Define an auxiliary two-vertex ribbon graph $\Gamma$ with loops $a,b$ and directed edges $C,D$ from $u$ to $v$, endowed with the cyclic orders \eqref{eq:hopf-active-orders} and joint magnetic projector $P_\Gamma$. Forming the tensor product of $J_3$ on $a$, $J_3$ on $b$, $J_2$ on $C$, $J_1$ on $D$, and $|0\rangle$ on all grey edges yields an isometry $J$ into the torus Hilbert space obeying
\begin{equation}
 Q_{\rm aux}P=J P_\Gamma J^\dagger.
 \label{eq:hopf-compression-identity}
\end{equation}
Appendix~\ref{subsec:hopf-path-isometry} establishes \eqref{eq:hopf-compression-identity} by verifying that $J$ carries $\operatorname{ran}P_\Gamma$ onto $Q_{\rm aux}\mathcal H_0$. The graph $\Gamma$ thus parametrizes the subspace $Q_{\rm aux}\mathcal H_0$ inside the $3\times3$ square torus while leaving the underlying lattice and regional edge supports intact.

Consider now an arbitrary preserving lift $\widehat A_{R_2}=A_{R_2}\otimes I_{R_2^c}\in K_{R_2}$ on the torus. Having support inside $R_2^c$, the projector $Q_{\rm aux}$ satisfies $[\widehat A_{R_2}\vert_{\mathcal H_0},Q_{\rm aux}\vert_{\mathcal H_0}]=0$; the physical adjoint of $\widehat A_{R_2}\vert_{\mathcal H_0}$ commutes with $Q_{\rm aux}\vert_{\mathcal H_0}$ as well, so both operators reduce $Q_{\rm aux}\mathcal H_0$. Denote by $U=J\vert_{\operatorname{ran}P_\Gamma}$ the unitary identification of $\operatorname{ran}P_\Gamma$ with $Q_{\rm aux}\mathcal H_0$, and write $J_b=J_3$ for the path isometry on $b=R_2$. Every other factor of $J$ acts on $R_2^c$, so conjugating $\widehat A_{R_2}\vert_{\mathcal H_0}$ by $U$ yields the restriction to $\operatorname{ran}P_\Gamma$ of the single-loop operator $A_b\otimes I_{\{a,C,D\}}$, with
\begin{equation}
 A_b=J_b^\dagger A_{R_2} J_b,
 \label{eq:hopf-row-lift}
\end{equation}
and the identity acting on $a,C,D$. Surjectivity of $U$ onto $Q_{\rm aux}\mathcal H_0$ guarantees that $A_b\otimes I_{\{a,C,D\}}$ leaves all of $\operatorname{ran}P_\Gamma$ invariant. Furthermore, on the physical operators on $\mathcal H_0$ that reduce $Q_{\rm aux}\mathcal H_0$, the map $X\mapsto U^\dagger XU$ is a star homomorphism. The two-way block separation established below is preserved under products and physical adjoints in the compressed representation of $\mathcal A(R_2)$; no multiplicativity is assumed for compression of arbitrary kinematic operators.

We restrict $A_b$ by testing it in the sector of $\Gamma$ where both $C$ and $D$ carry label $V$. As computed in Appendix~\ref{subsec:hopf-zero-blocks}, assigning a one-dimensional label $r\in\{0,\eta,s,\bar s\}$ to $b$ leaves a one-dimensional invariant space at $v$ spanned by an even Bell state of the $C,D$ endpoints, whereas assigning $V$ to $b$ produces a four-dimensional invariant space at $v$ with basis \eqref{eq:h8-four-invariants}. Any nonzero column of $A_b$ from a scalar label to $V$ must map that even Bell line into this four-dimensional space: parity forces the two odd entries to vanish, and comparing the two even basis vectors eliminates the remaining two entries. Dually, applying a row of $A_b$ from $V$ to a scalar label across the four basis states \eqref{eq:h8-four-invariants} forces all four row entries to vanish. Each local invariant state at $v$ used here extends to $\operatorname{ran}P_\Gamma$ by placing the input loop label on $a$ at $u$, so every preserving lift $A_b$ has zero off-diagonal blocks in both directions between $\{0,\eta,s,\bar s\}$ and $V$.

Block-diagonal separation between $\{0,\eta,s,\bar s\}$ and $V$ is closed under products and adjoints on $\operatorname{ran}P_\Gamma$, and the star homomorphism $X\mapsto U^\dagger XU$ transfers that separation to the action of $\mathcal A(R_2)$ on $Q_{\rm aux}\mathcal H_0$. Let $\Pi_V$ be the orthogonal projection on $\mathcal H_0$ onto states in which all three edges of $R_2$ carry label $V$. Observing that $\Psi_0\in Q_{\rm aux}\mathcal H_0$ carries label $0$ along $b$, we arrive at
\begin{equation}
 \Pi_V X_2\Psi_0=0
 \quad\text{for every }X_2\in\mathcal A(R_2).
 \label{eq:hopf-row-obstruction}
\end{equation}
Establishing \eqref{eq:hopf-row-obstruction} requires only the compressed representation on $Q_{\rm aux}\mathcal H_0$, with no claim that $\Pi_V$ commutes with $\mathcal A(R_2)$ on the full physical space.

\subsection{The entire generated algebra}
\label{subsec:hopf-generated-algebra}

Passing from \eqref{eq:hopf-row-obstruction} to general words in $\mathcal A(R_1)$ and $\mathcal A(R_2)$ rests on showing that the two regional star algebras commute on $\mathcal H_0$. Fix the six representation labels $\xi=(r_\epsilon)_{\epsilon\in E}$ on the connecting edges and denote the corresponding kinematic sector by $\mathcal H(\xi)$. Each edge $\epsilon\in E$ has one endpoint at each region. Let $W_{\epsilon,i}$ be $V_{r_\epsilon}$ if the half-edge at $R_i$ is outgoing and $V_{r_\epsilon}^{\vee}$ if it is incoming. Thus $F_i(\xi)=\bigotimes_{\epsilon\in E}W_{\epsilon,i}$ collects the six endpoint spaces on side $i$. Evaluate the ordered coproduct at each vertex starting at its cilium before regrouping the endpoints by region. This regrouping splits $\mathcal H(\xi)$ into two factors; each vertex projector acts on only one of them, so the restriction $P(\xi)$ of $P$ also factorizes:
\begin{equation}
 \mathcal H(\xi)
 =\bigl(\mathcal H_{R_1}\otimes F_1(\xi)\bigr)
  \otimes\bigl(\mathcal H_{R_2}\otimes F_2(\xi)\bigr),
 \qquad
 P(\xi)=P_1(\xi)\otimes P_2(\xi).
 \label{eq:hopf-endpoint-factorization}
\end{equation}
Here $P_1(\xi)$ multiplies the six vertex projectors in rows one and two, while $P_2(\xi)$ multiplies the three vertex projectors in row zero. Each preserving lift $\widehat A_{R_i}\in K_{R_i}$ leaves $\xi$ unchanged and acts trivially on the complementary factor in \eqref{eq:hopf-endpoint-factorization}. In every active sector with $P(\xi)\neq0$, pairing an arbitrary vector in $\operatorname{ran}P_i(\xi)$ with a fixed nonzero vector in $\operatorname{ran}P_{3-i}(\xi)$ produces a physical vector; invariance of the second factor under the identity then forces $\widehat A_{R_i}$ to preserve $\operatorname{ran}P_i(\xi)$ separately. Within $P(\xi)\mathcal H(\xi)$, elements of $\mathcal A_0(R_1)$ and $\mathcal A_0(R_2)$ therefore act as $X_1(\xi)\otimes I$ and $I\otimes X_2(\xi)$, forms inherited by their physical adjoints on $P(\xi)\mathcal H(\xi)$, whereas sectors with $P(\xi)=0$ contribute nothing to $\mathcal H_0$.

Operators in $\mathcal A(R_1)$ and $\mathcal A(R_2)$ consequently commute on $\mathcal H_0$, and finite-dimensionality of $\mathcal H_0$ reduces the generated star algebra $\mathcal A(R_1)\vee\mathcal A(R_2)$ to the linear span of products $X_1X_2$ with $X_i\in\mathcal A(R_i)$. Moreover, any lift in $K_{R_1}$ acts as the identity on $R_2$, ensuring that both its restriction to $\mathcal H_0$ and its physical adjoint commute with $\Pi_V$. Using $\Pi_V\Psi_V=\Psi_V$ alongside $[X_1,\Pi_V]=0$ and \eqref{eq:hopf-row-obstruction} yields
\begin{equation}
 \langle\Psi_V|X_1X_2|\Psi_0\rangle
 =\langle\Psi_V|X_1\Pi_VX_2|\Psi_0\rangle=0.
 \label{eq:hopf-generated-vanishing}
\end{equation}
Linearity extends this identity to $\langle\Psi_V|C|\Psi_0\rangle=0$ for every $C\in\mathcal A(R_1)\vee\mathcal A(R_2)$. Together with $\langle\Psi_V|W_{H_8}|\Psi_0\rangle=1$, the separating-matrix-element criterion in Section~\ref{subsec:separation-criterion} establishes Theorem~\ref{thm:hopf-obstruction}. Since $[W_{H_8},P]=0$, the obstruction remains valid if regional observables are restricted to reducing lifts commuting with $P$, and it requires neither a complete classification of $K_{R_2}$ nor star-closure of $\mathcal A_0(R_2)$.

\section{Discussion}
\label{sec:discussion}

In the disk example the same loop and two equidistant 24-edge supports give opposite answers for a finite non-Abelian group. Changing the contacts changes which exterior completions test the joint transition, not the magnetic physical space. What can the union measure in the nonadditive arrangement?

\subsection{What the union can detect}

The orthogonal relative-sign states $\psi_\pm$ of Corollary~\ref{cor:relative-sign} give identical probabilities for every effect in $\mathcal A(R_1)\vee\mathcal A(R_2)$, including effects sensitive to correlations between regions. The union witness has eigenvalues $+1$ and $-1$ on these states: one measurement determines the sign. Its operator-norm distance from the whole join is one. This algebra includes products and physical adjoints on $\mathcal H_0$, whether or not they have supported preserving lifts.

The wrapping pair admits a complete classification (Section~\ref{sec:wrapping-classification}). Fix the labels on the edges connecting the regions. In each active sector the join contains the orbit projections and every matrix unit within each orbit of row configurations under the group center. For non-Abelian $G$, the union algebra also connects different orbits within that same sector; neither algebra connects distinct connector sectors. The disk criterion (Theorem~\ref{thm:local-criterion}) decides additivity without classifying the blocks.

An algebra of effects specifies probabilities, not instruments or multi-round protocols. On a complete tensor product, unrestricted local operations and classical communication (LOCC) distinguish any two orthogonal pure states~\cite{WalgateShortHardyVedral2000LocalDistinguishability}. Mixed-state LOCC data hiding~\cite{TerhalDiVincenzoLeung2001Hiding} and local particle-number restrictions on accessible entanglement~\cite{WisemanVaccaro2003ParticleEntanglement} and state conversion~\cite{SchuchVerstraeteCirac2004SSRResources} concern different resources. Algebraic results on secret sharing~\cite{FiedlerNaaijkensOsborne2017Index}, purification~\cite{VanLuijkEtAl2026PurificationHaag} and entropic order~\cite{LiuVanLuijk2026EntropicHaag} require additional state or representation assumptions. Equal probabilities here imply neither an LOCC protocol bound nor a channel capacity.

For the non-diagonal $H_8$ constraints on the regular $3\times3$ torus, the union witness exchanges the physical states $\Psi_0$ and $\Psi_V$, with trivial and two-dimensional loop labels, whereas every operator in the join has zero matrix element between them (Section~\ref{sec:hopf-counterexample}). This is an independent obstruction, not a Hopf block classification.

\subsection{What restores additivity}

Ruling out one witness would not recover the full union algebra on the original supports. In the contact family, an uncovered inner face lets an exterior cusp label range over $G$ at fixed joint input. Only a uniform central shift of the loop passes every test. The peripheral factor preserves $\mathcal H_0$ on its own because its edges are contiguous in cyclic order at each boundary vertex (Lemma~\ref{lem:cuspless-factor}).

When $R$ avoids the testing neighborhood $D$ of the same loop (Theorem~\ref{thm:local-buffer}), the cusp tests still constrain the loop, but $R$ itself may have cusps. Its transition is initially known to preserve $\mathcal H_0$ only at one loop input. A unitary in $D$ changes the input without touching $R$, so preservation holds at every input (Eq.~\eqref{eq:transport-factor}). Thus in both geometries every preserving joint matrix unit with nonzero physical action has individually preserving factors. The regional algebras therefore generate the full union algebra on the prescribed supports. Neither result gives a distance criterion for arbitrary regions.

\subsection{Which regional algebras are being compared?}

The relabelings above act within the magnetic space. Harlow et al.\ instead take the full invariant space of a faithful, strongly continuous unitary action of a compact Lie group. Spatial factorization lets Haar averaging preserve support. Their additivity theorem also assumes commuting subgroups, each acting trivially on the other region, that generate the group together with the subgroup acting on neither~\cite{HarlowShaoSorceSrivastava2025Disjoint}. Group-valued edge labels do not establish these hypotheses for magnetic projectors. In a compatible $C^*$-module algebra, averaging over a finite-dimensional weak Hopf algebra gives a conditional expectation onto invariants after normalization if the identity averages to a finite nonzero scalar multiple of itself~\cite{AliAhmadKlingerWang2026ManyFaces}. Such an expectation need not provide representatives on prescribed edges. Cuspless additivity and collar inclusions~\cite{HolfesterSorce2026Gauss} are consistent with our counterexamples: the supports have cusps and no added collar.

Regional entropy depends on the subalgebra and its center~\cite{CasiniHuertaRosabal2014GaugeEntropy}, or on the embedding in an extended Hilbert space~\cite{Donnelly2012GaugeEntropy,GhoshSoniTrivedi2015GaugeEntropy,AokiEtAl2015GaugeEntropy}. For non-Abelian groups the latter adds a boundary representation-dimension term absent from electric-center entropy~\cite{Donnelly2014NonabelianEntropy}. Normal electric flux sectors contribute to the vacuum entropy of brick-wall-regulated free Maxwell theory~\cite{DonnellyWall2015ElectromagneticEdges}. Ground-state studies of generalized $\mathbb Z_2$ Fradkin--Shenker models compare entanglement spectra for exact and emergent Gauss laws and one-form symmetries~\cite{XuEtAl2025HigherFormEntanglement}. Operator-algebra quantum error correction asks which logical operators and their adjoints have subsystem representatives on a chosen code subspace~\cite{Harlow2017RTQEC}.

Boundary-extended regional phase spaces can be joined by symplectic reduction at the interface~\cite{DonnellyFreidel2016LocalSubsystems}. For a trivial gauge bundle over a simply connected sphere split into two balls, Gomes and Riello reconstruct the global reduced symplectic form when the bulk, regional and interface potentials are irreducible and the regional radiative data admit a continuously differentiable gluing. An interface mismatch can make the global form differ from the sum of the regional forms without independent boundary variables~\cite{GomesRiello2021QuasilocalYM,Riello2021EdgeModesWithout}. Lattice fusion bases use bulk excitation labels and boundary frames~\cite{DelcampDittrichRiello2016Entropy,DelcampDittrichRiello2017Fusion}; relational dressings use quantum reference frames and Wilson lines without adding kinematic degrees of freedom~\cite{AraujoRegadoHoehnSartini2025Relational}.

A maximal regional algebra can dualize to a nonmaximal one, so entropy comparisons must match the algebra and its trace~\cite{Radicevic2016EntropyDuality}; in the Ising chain the replica geometry selects the paired algebras~\cite{BulgarelliPanero2024ReplicaDuality}. Bulk correspondences between quantum doubles, string nets and topological field theories alone do not identify the regional algebras on these fixed edges~\cite{BuerschaperAguado2009Mapping,BalsamKirillov2012KitaevTV,Meusburger2017KitaevGauge}. For finite groups, gauging maps the equivariant rough-cut algebra of the trivial symmetry-protected model onto the smooth quantum-double boundary algebra. For the reversed cuts, only a distinguished equivariant smooth-cut subalgebra maps onto the rough boundary algebra: the omitted endpoint right actions have gauged images outside the patch~\cite{ChuahEtAl2024QuantumDoubleBoundary}. On an infinite chain a physical-boundary subalgebra determines a fusion category and is recovered as the fixed points of its fusion ring acting by bounded-spread unital completely positive maps~\cite{EvansJones2025FusionCategoryLattice}.

In quantum field theory, Haag duality identifies the regional algebra with the commutant of the algebra of its causal complement. Extended operators and generalized symmetries link this question to additivity without equating the two~\cite{CasiniEtAl2021EntropicOrder,CasiniMagan2021Completeness,GKSW2015}. Restricting states to the globally symmetric sector can break disjoint additivity~\cite{ShaoSorceSrivastava2025}. Symmetry topological field theory instead studies the regional algebra of invariant \emph{operators}. Its criteria for failures of additivity and duality specify the extended operators and assume every allowed symmetry representation occurs in the physical spectrum~\cite{JiaTian2026SymTFTAlgebras} (Table~\ref{tab:regional-prescriptions}).

\begin{table}[!htbp]
\centering
\caption{Comparison of regional composition frameworks across gauge, field-theoretic, and lattice settings. The retained regional data represent the mathematical inputs defining each problem rather than equivalent descriptions of a single underlying object.}
\label{tab:regional-prescriptions}
\small
\begin{tabular}{@{}p{.23\linewidth}p{.35\linewidth}p{.32\linewidth}@{}}
\hline
Setting & Regional data & Question \\
\hline
Gauge-field entropy \cite{CasiniHuertaRosabal2014GaugeEntropy,Donnelly2012GaugeEntropy,DelcampDittrichRiello2016Entropy} & Choice of regional observable center or extended boundary Hilbert space & Which regional reduced state and algebraic entropy are assigned to a subregion? \\
Regional phase space \cite{DonnellyFreidel2016LocalSubsystems,GomesRiello2021QuasilocalYM} & Boundary phase-space variables or compatible regional radiative data & How does symplectic reduction compose across an entangling interface? \\
Relational subsystems \cite{AraujoRegadoHoehnSartini2025Relational} & Chosen quantum reference frame and associated Wilson-line data & Which gauge-invariant subsystem algebra and, when available, physical factorization does the frame define? \\
Global symmetry sector~\cite{ShaoSorceSrivastava2025} & States restricted to the globally symmetric sector & Does disjoint additivity survive the restriction of the state space? \\
Symmetry-invariant observables~\cite{JiaTian2026SymTFTAlgebras} & Regional operators in the symmetry commutant, with specified extended operators & When does the invariant algebra satisfy additivity or Haag duality? \\
Infinite lattice nets \cite{Naaijkens2012ConeDuality,NaaijkensPenneysWallick2026LTOHaag,WallickWilming2026NonabelianHaag} & Cone von Neumann algebras in a specified ground-state or charged sector & Does the spatial complement algebra coincide with the commutant in the given representation? \\
Present finite model & The full magnetic physical subspace $\mathcal H_0$ and original edge supports & Do the two regional physical star algebras generate the physical algebra of their union? \\
\hline
\end{tabular}
\end{table}

In vacuum Gelfand--Naimark--Segal representations of the toric code and finite Abelian doubles, cones satisfy Haag duality and suitably separated nested cones satisfy distal split~\cite{Naaijkens2012ConeDuality,FiedlerNaaijkens2015AbelianDuality}. A split inclusion requires an intermediate type-I factor~\cite{DoplicherLongo1984Split}, not equality with a regional join. In an irreducible reference representation with cone duality and distal split, a two-cone index bounds the number of inequivalent irreducible transportable cone-localized charge sectors~\cite{Naaijkens2013ConeIndex}. For two-dimensional weak-Hopf-injective fixed-point states with biconnected $C^*$-weak Hopf input, the ground-state representation satisfies approximate cone duality on the original lattice and exact duality on adapted cones or, after $2\times2$ blocking, on standard cones~\cite{OgataEtAl2025HaagSpin}. In the local-topological-order framework, the boundary compatibility axiom for a cone and its complement (LTO-HD) gives cone duality in the canonical ground state under central-support and faithfulness assumptions~\cite{NaaijkensPenneysWallick2026LTOHaag}. Both exact and approximate cone duality can fail in a Levin--Wen sector carrying a single non-Abelian anyon~\cite{WallickWilming2026NonabelianHaag}. Here the representation is on the full magnetic space $\mathcal H_0$.

{\samepage Which other geometries allow exterior labels to vary at fixed joint input and loop inputs to change without touching the companion? A criterion for general Hopf constraints would need tests on invariant tensors rather than classical labels. A continuum statement would also require a choice of representation and scaling limit.\par}

\appendix
\section{Details of the wrapping algebras}
\label{app:wrapping-details}
\label{app:diagonal-fibres}

Here we establish the represented block decomposition of Theorem~\ref{thm:wrapping-blocks} for the wrapping partition of Section~\ref{sec:wrapping-classification} ($N,M\ge3$) by applying the exterior-fibre criterion of Section~\ref{subsec:fibre-criterion} on the full magnetic subspace $\mathcal H_0$.

\textit{Extensions, fibres, and sector sizes.}---Evaluating the vertex constraint products directly confirms that the row extension in Eq.~\eqref{eq:arbitrary-row-extension} is physically valid. Along rows $j=0$ and $j=1$, the counterclockwise vertex products evaluate to $b_i c_i b_{i-1}^{-1}d_i^{-1}=e$ and $b_i^{-1}d_i b_{i-1}c_i^{-1}=e$, respectively. At every remaining row $2\leq j\leq M-1$, the constraint reduces to $d_i d_i^{-1}=e$, which holds down to the minimal height $M=3$. Every row tuple $b\in G^N$ and connector assignment $c\in G^N$ therefore admits at least one physical completion.

To apply the exterior-fibre criterion of Section~\ref{subsec:fibre-criterion}, we determine the active connector sectors and the resulting matrix blocks on $\mathcal H_0$. For any $R_1$ configuration $r$ that satisfies all interior vertex constraints at $2\leq j\leq M-2$ (a condition that is vacuous when $M=3$), the boundary Gauss laws along rows $1$ and $M-1$ fix the adjacent vertical connector tuples $\xi=(c(r),d(r))$ via
\begin{align}
 c_i(r)&=h_{i,1}v_{i,1}h_{i-1,1}^{-1},\nonumber\\
 d_i(r)&=h_{i,M-1}^{-1}v_{i,M-2}h_{i-1,M-1}.
 \label{eq:em-connectors}
\end{align}
For $r\in X_\xi$, the allowed exterior fibre on $R_2\cup E$ is $Y_\xi\times\{\xi\}$. For a physical union configuration $(r,b)$ on $R_1\cup R_2$ with $r\in X_\xi$ and $b\in Y_\xi$, the exterior fibre on $E$ reduces to the singleton $\{\xi\}$. In both cases, configurations with equal connector data share the same nonempty fibre, whereas distinct connector assignments yield disjoint fibres.

Set $D_\xi=d_{N-1}\cdots d_0$. Propagating the row compatibility relation $b_i=d_i b_{i-1}c_i^{-1}$ around the cycle from the initial entry $s=b_{N-1}$ closes consistently if and only if
\begin{equation}
 s=D_\xi s C_\xi^{-1},
 \qquad D_\xi=sC_\xi s^{-1}.
 \label{eq:em-sector-conjugacy}
\end{equation}
Accordingly, $Y_\xi$ is nonempty if and only if the holonomies $C_\xi$ and $D_\xi$ are conjugate in $G$; when this holds, the valid initial elements $s$ form a single coset of the centralizer $C_G(C_\xi)$ of cardinality $|Y_\xi|=\kappa_\xi$. Any $b\in Y_\xi$ guarantees that $X_\xi$ is nonempty via the explicit completion in Eq.~\eqref{eq:arbitrary-row-extension}, so $\xi=(c,d)$ belongs to the active set $\Xi$ precisely when $C_\xi$ and $D_\xi$ are conjugate. Conversely, iterating the interior vertex relation $v_{i,j}=h_{i,j}^{-1}v_{i,j-1}h_{i-1,j}$ across the rows of $R_1$ gives $d_i=a_i^{-1}c_i a_{i-1}$ with vertical product $a_i=h_{i,1}\cdots h_{i,M-1}$. Setting $b_i=a_i^{-1}$ produces a tuple $b\in Y_\xi$. For each $b\in Y_\xi$, the first $M-2$ horizontal rows of $R_1$ may be assigned freely in $G^N$, the top row $h_{i,M-1}$ is fixed uniquely by $a_i=b_i^{-1}$, and all interior vertical edges $v_{i,j}$ ($1\le j\le M-2$) are determined recursively from $c_i$. This bijection between $X_\xi$ and $G^{N(M-2)}\times Y_\xi$ yields $m_\xi=|X_\xi|=|G|^{N(M-2)}\kappa_\xi$.

\textit{Matrix units and independent sector coefficients.}---Because nonempty fibres for $R_1$ and for $R_1\cup R_2$ are pairwise identical or disjoint, the criterion of Section~\ref{subsec:fibre-criterion} shows that their raw physical images are star-closed and decompose as
\begin{align}
 \mathcal A_0(R_1)&=\bigoplus_{\xi\in\Xi}
       \operatorname{End}(\mathbb C[X_\xi])\otimes I_{Y_\xi},\nonumber\\
 \mathcal A_0(R_1\cup R_2)&=\bigoplus_{\xi\in\Xi}
       \operatorname{End}(\mathbb C[X_\xi]\otimes\mathbb C[Y_\xi]).
 \label{eq:em-raw-blocks}
\end{align}
Kinematic lifts in $K_R$ may also carry arbitrary columns on configurations with empty exterior fibres, which vanish upon restriction to $\mathcal H_0$; Eq.~\eqref{eq:em-raw-blocks} records the restricted physical algebras on $\mathcal H_0$ rather than the kinematic spaces $K_R$. On the second region, Eq.~\eqref{eq:row-central-shift} shows that $\mathcal A_0(R_2)$ is a star-closed faithful representation of $\bigoplus_{O\in G^N/Z(G)}\operatorname{End}(\mathbb C[O])$, where each orbit matrix unit acts with identical coefficients across every active sector $\xi\in\Xi$ such that $O\subseteq Y_\xi$. Within each sector $\xi$, $\mathcal A(R_1)$ and $\mathcal A(R_2)$ act on the separate tensor factors $\mathbb C[X_\xi]$ and $\mathbb C[Y_\xi]$ and therefore commute.

Because both regional algebras preserve the connector label $\xi$ and the diagonal center orbit $O\in Y_\xi/Z(G)$, the generated algebra $\mathcal B=\mathcal A(R_1)\vee\mathcal A(R_2)$ lies inside the block-diagonal algebra of Eq.~\eqref{eq:complete-blocks}. For the reverse inclusion, define the $R_1$-supported sector projector
\begin{equation}
 \widehat{\Pi}_\xi=
 \left(\sum_{r\in X_\xi}|r\rangle\langle r|\right)_{R_1}\otimes I,
 \qquad \widehat{\Pi}_\xi|_{\mathcal H_0}=\Pi_\xi .
 \label{eq:em-sector-selector}
\end{equation}
The operator $\widehat{\Pi}_\xi$ commutes with $P$ and restricts on $\mathcal H_0$ to the orthogonal projector $\Pi_\xi$ onto sector $\xi$. For any $r,r'\in X_\xi$ and any $b,b'\in Y_\xi$ lying in the same center orbit $O$, the restricted product of the corresponding regional matrix units gives the elementary sector matrix unit
\begin{equation}
 \left.(E^{R_1}_{r',r}E^{R_2}_{b',b})\right|_{\mathcal H_0}
       =|r',b',\xi\rangle\langle r,b,\xi|.
 \label{eq:em-block-matrix-unit}
\end{equation}
Multiplying by $\Pi_\xi$ decouples the shared $R_2$ orbit coefficients across distinct connector sectors, so linear combinations of the products in Eq.~\eqref{eq:em-block-matrix-unit} span each matrix block $M_{m_\xi z_G}(\mathbb C)$ independently. This completes the proof of Theorem~\ref{thm:wrapping-blocks}.

Finally, on the witness subspace $Q\mathcal H_0$ defined by setting $r=\mathbf e_{R_1}$, the boundary relations select the trivial connector sector $\xi=(e,e)$ and restrict $Y_{(e,e)}$ to the uniform rows $\bar g$, specializing Eq.~\eqref{eq:complete-blocks} to the corner decomposition in Eq.~\eqref{eq:witness-corners}. Whenever $G$ is nontrivial, $m_{(e,e)}=|G|^{N(M-2)+1}>1$, so we may choose $r_1\in X_{(e,e)}$ with $r_1\neq\mathbf e_{R_1}$. The $R_1$ lift $|r_1\rangle\langle\mathbf e_{R_1}|\otimes I$ preserves $\mathcal H_0$ and restricts to a partial isometry $V\in\mathcal A(R_1)\subseteq\mathcal B$ satisfying $QVQ=0$ whereas $QV^\dagger VQ=Q$. Compression by $Q$ is therefore not a star homomorphism on $\mathcal B$. Instead, $Q\mathcal BQ$ and $Q\mathcal A_UQ$ form unital associative algebras because $Q$ itself lies in $\mathcal B$, and for any physical state $\rho$ supported on $Q\mathcal H_0$ ($\rho=Q\rho Q$), cycling $Q$ inside the trace gives $\operatorname{Tr}(\rho X)=\operatorname{Tr}(\rho QXQ)$, which preserves their operational role in governing measurement statistics on $Q\mathcal H_0$.

\section{One-sided locking and a buffered wrapping strip}
\label{sec:wrapping-variants}
Modifying the wrapping geometry provides two complementary probes of boundary contact on the full magnetic subspace, using the stated regional supports for each geometry. Shrinking the first region to leave direct adjacency along only one boundary still produces a counterexample for every finite non-Abelian group, whereas the buffered-strip construction with exterior buffer rows on both sides restores additivity for all finite groups and yields an explicit block decomposition.

\subsection{A one-sided counterexample}
Consider a torus of dimensions $N\ge3$ and $M\ge4$. Retaining the single-row subsystem $R_2=\{h_{i,0}\}$, we replace the first region $R_1$ with
\begin{equation}
 S_- =\{h_{i,M-1},v_{i,M-2}:i\in\mathbb Z_N\}.
 \label{eq:one-sided-region}
\end{equation}
The vertices of $S_-$ are confined to rows $M-2$ and $M-1$, ensuring that $V(S_-)\cap V(R_2)=\emptyset$. Consequently, $S_-$ directly neighbors row zero exclusively through the southern vertical boundary edges $d_i=v_{i,M-1}$; the northern row one contains no vertices of $S_-$. This one-sided contact structure persists down to the minimal height $M=4$, where row zero reaches $S_-$ in two vertical steps through the northern row one.

Let $Q_-$ be the diagonal kinematic projector enforcing the identity element on all edges of $S_-$; it commutes directly with the full magnetic projector $P$. On any physical basis state within the support of $Q_-$, the vertex constraint at $(i,M-1)$ uniquely fixes $d_i=e$. With $b_i=h_{i,0}$ and $c_i=v_{i,0}$, the row-zero vertex constraint then reduces to
\begin{equation}
 b_i c_i b_{i-1}^{-1}=d_i=e.
 \label{eq:one-sided-lock}
\end{equation}
For any non-Abelian group $G$, select a non-central element $t\notin Z(G)$ and form the row-exchange operator $T_{R_2}$ defined in Eq.~\eqref{eq:group-joint-witness}. The composite operator $\widehat W_-=Q_-T_{R_2}$ annihilates all basis states except those satisfying $Q_-=1$ whose row zero evaluates to either uniform configuration $\bar e$ or $\bar t$. On these active physical inputs, Eq.~\eqref{eq:one-sided-lock} forces the intermediate vertical edges to $c_i=e$. Interchanging $\bar e$ and $\bar t$ on row zero therefore leaves every row-zero vertex constraint satisfied while leaving all other edge labels and vertex words untouched. Hence $\widehat W_-$ preserves the physical Hilbert space $\mathcal H_0$, and its Hermiticity ensures $[\widehat W_-,P]=0$.

The witness states $\psi_e,\psi_t$ satisfy $Q_-\psi_g=\psi_g$ and yield the transition amplitude $\langle\psi_t|\widehat W_-|\psi_e\rangle=1$. Because $S_-$ is a strict subset of $R_1$ in Eq.~\eqref{eq:region-partition}, support monotonicity imposes
\begin{equation}
 \mathcal A(S_-)\vee\mathcal A(R_2)
 \subseteq\mathcal A(R_1)\vee\mathcal A(R_2).
 \label{eq:one-sided-inclusion}
\end{equation}
The vanishing matrix element $\langle\psi_t|C|\psi_e\rangle=0$ established in Eq.~\eqref{eq:group-separation} for all $C\in\mathcal A(R_1)\vee\mathcal A(R_2)$ therefore restricts directly to $\mathcal A(S_-)\vee\mathcal A(R_2)$. The kinematic bound $\|\widehat W_-\|\le\|Q_-\|\,\|T_{R_2}\|=1$ and the unit witness matrix element give $\|W_-\|=1$ for the physical restriction $W_-=\widehat W_-|_{\mathcal H_0}$. For every $C$ in the smaller join, the same matrix element yields $\|W_--C\|\ge1$, while $C=0$ attains the upper bound. Thus $\operatorname{dist}(W_-,\mathcal A(S_-)\vee\mathcal A(R_2))=1$. The mutual commutativity $[\mathcal A(S_-),\mathcal A(R_2)]=0$ is inherited immediately from the parent inclusion. This establishes additivity failure for every finite non-Abelian group $G$ at $N\ge3, M\ge4$. Direct locking from both boundaries is therefore unnecessary; however, this counterexample still exploits direct adjacency along one boundary, topological wrapping, and the cusp of $R_2$, without asserting an Abelian converse for this subregion.

\subsection{The complete buffered-strip algebra}
To isolate the effect of separating the boundaries, let $N\ge3$, $M\ge5$, keep $R_2=\{h_{i,0}\}$, and define the buffered strip
\begin{equation}
 S_b=\{h_{i,j}:2\le j\le M-2\}
       \cup\{v_{i,j}:2\le j\le M-3\}.
 \label{eq:buffered-strip-region}
\end{equation}
Its vertices span rows $2,\ldots,M-2$, leaving rows $1$ and $M-1$ as exterior buffers. The graph distance between $V(S_b)$ and $V(R_2)$ is therefore exactly two on the torus. The restriction $M\ge5$ ensures that the strip contains internal vertical edges from which both boundary tuples can be reconstructed. At $M=4$, the two buffer rows would still be distinct, but $S_b$ would contain only one horizontal row, so the boundary reconstruction used below would not apply.

Setting $H=G^N$, $q=|G|$, and $z=|Z(G)|$, we define the right group action
\begin{equation}
 \alpha_x(u)_i=x_i^{-1}u_i x_{i-1},\qquad
 \alpha_y\!\left(\alpha_x(u)\right)=\alpha_{xy}(u).
 \label{eq:strip-action}
\end{equation}
Let the row tuples be $x_j=(h_{i,j})_i$ and $u_j=(v_{i,j})_i$, so that the magnetic vertex condition translates into the recursion $u_j=\alpha_{x_j}(u_{j-1})$. We label the key boundary and buffer variables by
\begin{equation}
 b=x_0,\quad a=x_1,\quad f=x_{M-1},\quad
 c=u_0,\quad\ell=u_1,\quad k=u_{M-2},\quad d=u_{M-1}.
 \label{eq:strip-variables}
\end{equation}
The tuple $(a,f,c,\ell,k,d)$ labels precisely the edges in $(S_b\cup R_2)^c$, the complement of the union rather than of the strip $S_b$ alone. Any admissible internal configuration $r$ on $S_b$ fixes the boundary tuple $\eta=(\ell,k)$ via its outermost vertex relations,
\begin{align}
 \ell_i(r)&=h_{i,2}v_{i,2}h_{i-1,2}^{-1},\nonumber\\
 k_i(r)&=h_{i,M-2}^{-1}v_{i,M-3}h_{i-1,M-2}.
 \label{eq:strip-boundaries}
\end{align}
The remaining vertex conditions on the buffer rows and row zero become
\begin{equation}
 \ell=\alpha_a(c),\qquad d=\alpha_f(k),\qquad c=\alpha_b(d).
 \label{eq:strip-exterior-equations}
\end{equation}
For a given boundary pair $\eta$, we denote by $X_\eta$ the sector of physically admissible configurations on $S_b$ compatible with $\eta$. Non-emptiness of $X_\eta$ is equivalent to conjugacy in $G$ of the ordered products $C_\ell=\ell_{N-1}\cdots\ell_0$ and $C_k=k_{N-1}\cdots k_0$; we therefore collect the active boundary pairs in the domain $\Xi_b=\{(\ell,k)\in H\times H:C_\ell\text{ and }C_k\text{ are conjugate in }G\}$. Fixing $\eta\in\Xi_b$ and writing $\kappa_\eta=|C_G(C_\ell)|$, we enumerate $X_\eta$ by freely specifying the first $M-4$ horizontal rows of $S_b$ in $q^{N(M-4)}$ ways while constraining the total horizontal product $A=x_2x_3\cdots x_{M-2}$ through $\alpha_A(\ell)=k$. Propagating the coordinate relation $A_{i-1}=\ell_i^{-1}A_i k_i$ around the torus expresses the entire tuple $A$ in terms of $A_{N-1}$, subject to the periodic closure condition $A_{N-1}^{-1}C_\ell A_{N-1}=C_k$. Since $G$ contains $\kappa_\eta$ such conjugating elements $A_{N-1}$, the relation $\alpha_A(\ell)=k$ admits $\kappa_\eta$ distinct solutions $A$. Together with the chosen preceding rows, each solution $A$ uniquely determines the last horizontal row $x_{M-2}$, from which the vertex recursion fixes every internal vertical edge of $S_b$. This one-to-one parametrization establishes the sector dimension
\begin{equation}
 m_\eta:=|X_\eta|=q^{N(M-4)}\kappa_\eta,\qquad \eta\in\Xi_b.
 \label{eq:strip-sector-size}
\end{equation}

For every active boundary pair $\eta$ and any row configuration $b\in H$, the joint physical exterior completion fiber is
\begin{equation}
 F_{\eta,b}=\{(a,f,c,\ell,k,d):d=\alpha_f(k),\ c=\alpha_b(d),\ \alpha_a(c)=\ell\}.
 \label{eq:strip-fiber}
\end{equation}
Because $f\in H$ is completely unconstrained, it has $q^N$ possible values; once $f$ is chosen, $d$ and $c$ are fixed within the $\alpha$-orbit of $k$, leaving exactly $\kappa_\eta$ choices of $a$ that satisfy $\alpha_a(c)=\ell$. The fiber size is therefore uniformly $|F_{\eta,b}|=q^N\kappa_\eta$. Crucially, as $f$ sweeps over $H$, each individual coordinate $d_i$ can attain any prescribed group element $x\in G$: setting $f_i=e$ and $f_{i-1}=k_i^{-1}x$ ensures $d_i=x$ regardless of the remaining coordinates. This establishes sufficient coordinate freedom within the orbit of $k$, without asserting unconstrained simultaneous specification of the entire tuple $d$.

For diagonal constraints, a supported kinematic matrix unit preserves $\mathcal H_0$ precisely when every exterior completion of its input is also a completion of its output. In an arbitrary lift, leakage between a fixed full input and a fixed full output has a single local matrix coefficient and therefore cannot cancel between distinct matrix units. Now consider a preserving unit $|r',b'\rangle\langle r,b|$ on $S_b\cup R_2$ with nonzero physical action. Because $\ell$ and $k$ are exterior edges that couple to the internal boundary equations, physical admissibility requires that the input configuration $r$ and output configuration $r'$ share the identical boundary pair $\eta$. Mapping physical input completions into physical output completions across the entire fiber further demands $\alpha_{b'}(d)=\alpha_b(d)$ for all $d$ in the $\alpha$-orbit of $k$. Setting $s_i=b'_i b_i^{-1}$, this invariance condition is equivalent to
\begin{equation}
 s_i d_i=d_i s_{i-1}\quad\text{for every allowed }d.
 \label{eq:strip-center-test}
\end{equation}
Evaluating Eq.~\eqref{eq:strip-center-test} at an allowed configuration with $d_i=e$ forces $s_i=s_{i-1}$, implying that $s_i$ is a uniform group element $s$ across all sites. Evaluating it at configurations where $d_i$ attains an arbitrary element $x\in G$ then requires $sx=xs$, so that $s=\zeta\in Z(G)$. Conversely, any uniform central shift $b'=b\zeta$ leaves $\alpha_b(d)$ invariant for every $d$. Hence, the non-zero physical matrix units on the union are precisely those satisfying $r,r'\in X_\eta$ for a common $\eta$ and $b'=b\zeta$ with $\zeta\in Z(G)$.

Every matrix unit $|r'\rangle\langle r|$ with $r,r'\in X_\eta$ defines an individually admissible physical lift on $S_b$, as its exterior compatibility condition depends solely on $\eta$; its adjoint $|r\rangle\langle r'|$ is likewise admissible. On $R_2$, the row shifts $|b\zeta\rangle\langle b|$ with $\zeta\in Z(G)$ are individually admissible by Eq.~\eqref{eq:row-central-shift}. Each allowed union matrix unit is the product of these two regional units. Expanding an arbitrary preserving union lift and discarding columns with empty input fibers, which act as zero on $\mathcal H_0$, therefore expresses its physical restriction as a linear combination of those products. The coefficients remain identical across all exterior completions of each input. Passing to the generated physical $C^*$-algebras establishes additivity,
\begin{equation}
 \mathcal A(S_b\cup R_2)=\mathcal A(S_b)\vee\mathcal A(R_2).
 \label{eq:strip-additivity}
\end{equation}
Because the raw physical representations in this geometry are closed under the adjoint, their disjoint-support lifts generate commuting physical $C^*$-algebras on $\mathcal H_0$.

This classification also explicitly reveals the block structure of the represented physical algebra and its external multiplicities. Letting $O$ index the orbits of the free diagonal action of $Z(G)$ on $H$, and denoting by $F_{\eta,O}$ the fiber shared by all $b\in O$, the Hilbert space and algebra decompose as
\begin{align}
 \mathcal H_0&\cong\bigoplus_{\substack{\eta\in\Xi_b\\O\in H/Z(G)}}
 \mathbb C[X_\eta]\otimes\mathbb C[O]\otimes\mathbb C[F_{\eta,O}],\nonumber\\
 \mathcal A(S_b\cup R_2)&\cong\bigoplus_{\substack{\eta\in\Xi_b\\O\in H/Z(G)}}
 M_{m_\eta z}(\mathbb C)\otimes I_{q^N\kappa_\eta}.
 \label{eq:strip-complete-blocks}
\end{align}
Sums of diagonal $S_b$ matrix units project onto individual $\eta$ sectors, and sums of diagonal $R_2$ matrix units select individual orbits $O$. Within each selected pair $(\eta,O)$, products of the regional matrix units span the full factor $M_{m_\eta z}(\mathbb C)$. The explicit identity factor $I_{q^N\kappa_\eta}$ reflects the fact that regional operator coefficients are shared identically across all exterior fiber completions. This positive additivity theorem applies specifically to the buffered cylindrical geometry; it complements the local-loop theorem of Section~\ref{sec:local-criterion} without asserting a general distance criterion for arbitrary subsystems.

\section{Hopf data and the supported reduction}
\label{app:hopf-data}

Here we fix the $H_8$ algebra conventions and establish the supported path-isometry reduction \eqref{eq:hopf-compression-identity} together with the vanishing of the scalar--$V$ blocks invoked in Section~\ref{sec:hopf-counterexample}. Throughout, we work within the magnetic-only quantum-double framework of \cite{HolfesterSorce2026Gauss} and adopt the explicit $H_8$ presentation of \cite{LuChatterjeeTantivasadakarn2026}.

\subsection{Representations and magnetic projectors}
\label{subsec:hopf-data}

The Hopf algebra $H_8$ is generated by $x,y,z$ subject to the relations and star structure
\begin{align}
 x^2&=y^2=1,& xy&=yx,& zx&=yz,& zy&=xz,\nonumber\\
 z^2&=\frac{1+x+y-xy}{2},&
 x^*&=x,& y^*&=y,& z^*&=z^3,
 \label{eq:h8-algebra-data}
\end{align}
with coproducts
\begin{align}
 \Delta x&=x\otimes x,&\Delta y&=y\otimes y,\nonumber\\
 \Delta z&=\frac12(1\otimes1+y\otimes1+1\otimes x-y\otimes x)
                         (z\otimes z).
 \label{eq:h8-coproduct-data}
\end{align}
The counit satisfies $\epsilon(x)=\epsilon(y)=\epsilon(z)=1$, and the antipode $S$ fixes $x,y,z$. The normalized Haar integral is
\begin{equation}
 \Lambda=\frac18(1+x+y+xy)(1+z).
 \label{eq:h8-integral}
\end{equation}
Here $(1+x)(1+y)/4$ is a central self-adjoint idempotent on whose range $z^2=1$ and $z^*=z$, so $\Lambda$ is a self-adjoint idempotent as well. Because $a\Lambda=\epsilon(a)\Lambda$ for all $a\in H_8$ and $\epsilon(\Lambda)=1$, $\Lambda$ projects any representation onto its subspace of invariant vectors.

We fix the five irreducible representations of $H_8$ as
\begin{equation}
 \begin{array}{c|ccc}
 r&\rho_r(x)&\rho_r(y)&\rho_r(z)\\ \hline
 0&1&1&1\\
 \eta&1&1&-1\\
 s&-1&-1&i\\
 \bar s&-1&-1&-i\\
 V&Z&-Z&X
 \end{array}
 \qquad
 Z=\begin{pmatrix}1&0\\0&-1\end{pmatrix},\quad
 X=\begin{pmatrix}0&1\\1&0\end{pmatrix}.
 \label{eq:h8-representations}
\end{equation}
An incoming half-edge carries the contragredient representation
\begin{equation}
 \rho_r^\vee(a)=\rho_r(S(a))^{\mathsf T}.
 \label{eq:h8-contragredient}
\end{equation}
Because $S$ fixes $x,y,z$ and the matrices in \eqref{eq:h8-representations} are symmetric, $\rho_r^\vee(a)=\rho_r(a)$ for every $r$ in this basis. In particular, the one-dimensional representations $s$ and $\bar s$ are each self-contragredient; taking the Hopf dual does not conjugate their scalar matrix entries.

We equip the edge algebra $H=K^\vee$ with the Haar inner product, under which the matrix coefficients $Z^r_{ij}$ satisfy
\begin{equation}
 \langle Z^r_{ij},Z^{r'}_{k\ell}\rangle
 =\frac{1}{d_r}\delta_{rr'}\delta_{ik}\delta_{j\ell}.
 \label{eq:h8-pw-normalization}
\end{equation}
Thus the endpoint-ordered basis $|r,i,j\rangle_{\rm end}=\sqrt{d_r}\,Z^r_{ji}$ of \eqref{eq:hopf-endpoint-basis} is orthonormal and lists the outgoing index $i$ first and the incoming index $j$ second. By contrast, the source basis $|r,i,j\rangle_Z=\sqrt{d_r}\,Z^r_{ij}$ lists the incoming index first; in those coordinates, the two-segment path isometry reads
\begin{equation}
 J_{2,Z}|r,i,j\rangle_Z
 =d_r^{-1/2}\sum_k
 |r,k,j\rangle_{Z,1}\otimes|r,i,k\rangle_{Z,2}.
 \label{eq:h8-source-path}
\end{equation}
Transposing the two indices within each edge converts \eqref{eq:h8-source-path} into the endpoint-ordered formula \eqref{eq:hopf-path-map} for $m=2$, with the path's edge tensor factors always ordered from source to target.

At a vertex $v$ of degree $d_v$, let $\Delta^{(d_v-1)}$ denote the iterated coproduct into $d_v$ factors. In a sector of fixed incident edge labels, the magnetic projector is given by
\begin{equation}
 P_v=(\rho_{v,1}\otimes\cdots\otimes\rho_{v,d_v})
                         \bigl(\Delta^{(d_v-1)}(\Lambda)\bigr),
 \label{eq:h8-vertex-projector}
\end{equation}
where $\rho_{v,k}$ is $\rho_r$ for an outgoing half-edge and $\rho_r^\vee$ for an incoming half-edge, ordered according to the vertex cilium. We always evaluate the ordered coproduct action \eqref{eq:h8-vertex-projector} before regrouping endpoint Hilbert spaces. Because distinct vertices act on disjoint sets of half-edges and leave the representation labels $r$ unchanged, all $P_v$ commute.

\subsection{The full path-isometry image}
\label{subsec:hopf-path-isometry}

By the counit property $(\operatorname{id}\otimes\epsilon)\circ\Delta=\operatorname{id}=(\epsilon\otimes\operatorname{id})\circ\Delta$, any half-edge fixed to the trivial representation $r=0$ drops out of the coproduct in \eqref{eq:h8-vertex-projector}. For a $K$-module $M$ equipped with action $\rho_M$, let $\operatorname{Inv}(M)=\{w\in M:\rho_M(a)w=\epsilon(a)w\ \text{for all }a\in K\}$ denote the subspace of invariants, where $\epsilon$ is the counit. At a vertex with one incoming half-edge of label $r$ and one outgoing half-edge of label $s$, Schur's lemma yields
\begin{equation}
 \dim\operatorname{Inv}(V_r^\vee\otimes V_s)=\delta_{rs}.
 \label{eq:h8-bivalent-dimension}
\end{equation}
For $r=s$ the unique invariant line is spanned by the normalized maximally entangled state $d_r^{-1/2}\sum_k|k\rangle\otimes|k\rangle$. Along an oriented $m$-edge path, imposing this bivalent singlet projector at each of the $m-1$ internal vertices forces all $m$ edges to carry the same label $r$ and contracts the internal endpoint indices according to \eqref{eq:hopf-path-map}. Orthonormality of the endpoint basis gives $J_m^\dagger J_m=I$, and the dimension count \eqref{eq:h8-bivalent-dimension} shows that $\operatorname{ran}J_m$ is the entire subspace satisfying the internal bivalent constraints. Moreover, $J_m$ intertwines the single-edge endpoint actions on $|r,i,j\rangle$ with the outer endpoint actions at the source of the first segment and the target of the $m$th segment.

On $\operatorname{ran}Q_{\rm aux}$, the active paths $a,b,C,D$ have lengths $3,3,2,1$ (Figure~\ref{fig:hopf-reduction}). Deleting the trivial half-edges fixed by $Q_{\rm aux}$ leaves $u$ and $v$ four-valent with the cyclic orders \eqref{eq:hopf-active-orders}, while each of the other seven vertices either becomes a bivalent internal vertex on $a$, $b$, or $C$ or has only trivial incident edges and satisfies $P_v=1$. By the bivalent path classification above, the subspace of $\operatorname{ran}Q_{\rm aux}$ satisfying all vertex projectors outside $\{u,v\}$ is precisely the image of the isometry $J$ obtained by tensoring $J_3$ on $a$, $J_3$ on $b$, $J_2$ on $C$, $J_1$ on $D$, and the trivial state on the grey edges. Because $J_m$ intertwines the outer endpoint actions on each path, $J$ intertwines the remaining vertex projectors $P_uP_v$ with $P_\Gamma$. It therefore maps $\operatorname{ran}P_\Gamma$ unitarily onto $Q_{\rm aux}\mathcal H_0$, which establishes \eqref{eq:hopf-compression-identity}.

Because the factor $J_b=J_3$ acts strictly on $b=R_2$ and all other factors of $J$ act on $R_2^c$, compressing any lift $\widehat A_{R_2}=A_{R_2}\otimes I_{R_2^c}$ by $J$ yields $J^\dagger\widehat A_{R_2}J=A_b\otimes I_{\{a,C,D\}}$ with $A_b=J_b^\dagger A_{R_2}J_b$. If $\widehat A_{R_2}\in K_{R_2}$, then $\widehat A_{R_2}$ preserves $Q_{\rm aux}\mathcal H_0=J\operatorname{ran}P_\Gamma$, so $A_b\otimes I_{\{a,C,D\}}$ preserves all of $\operatorname{ran}P_\Gamma$. And because $Q_{\rm aux}\vert_{\mathcal H_0}$ commutes with $\mathcal A_0(R_2)$ and hence with $\mathcal A(R_2)$, restricting to $Q_{\rm aux}\mathcal H_0$ and conjugating by $U=J\vert_{\operatorname{ran}P_\Gamma}$ defines a star homomorphism on $\mathcal A(R_2)$, without requiring $U$ to identify $\mathcal A(R_2)$ with the full single-edge algebra on $\Gamma$.

\subsection[Blocks between scalar and two-dimensional labels]{Scalar--\texorpdfstring{$V$}{V} blocks of a single-loop lift}
\label{subsec:hopf-zero-blocks}

It remains to compute the vertex projectors of $\Gamma$ when $C$ and $D$ carry label $V$ and a loop carries label $r$. At $u$, the four representation factors in cyclic order are $(r,V,r^\vee,V)$. After applying $\Delta^{(3)}(\Lambda)$ in that order, we permute the four half-edge factors so that the two loop endpoints precede the two exterior endpoints,
\begin{equation}
 \mathcal L_r\otimes\mathcal F_{VV}
 =(V_r\otimes V_r^\vee)\otimes(V\otimes V).
 \label{eq:h8-local-grouping}
\end{equation}
Denote the resulting projector on $\mathcal L_r\otimes\mathcal F_{VV}$ by $P_{r;VV}$. At $v$, the exterior half-edges are incoming, but since $\rho_V^\vee=\rho_V$ by \eqref{eq:h8-contragredient}, the regrouped projector at $v$ is given by the exact same matrix $P_{r;VV}$.

When $r\in\{0,\eta,s,\bar s\}$ is one-dimensional, $\mathcal L_r\simeq\mathbb C$ and substituting \eqref{eq:h8-representations} into \eqref{eq:h8-vertex-projector} gives
\begin{equation}
 P_{r;VV}=|\phi_{\sigma(r)}\rangle\langle\phi_{\sigma(r)}|,
 \qquad
 \phi_\sigma=\frac{|00\rangle+\sigma|11\rangle}{\sqrt2},
 \qquad
 \sigma(0)=\sigma(\eta)=1,\quad
 \sigma(s)=\sigma(\bar s)=-1.
 \label{eq:h8-scalar-bell-lines}
\end{equation}
Indeed, on $V\otimes V$ both $x$ and $y$ act as $Z\otimes Z$, so their invariance conditions retain $\operatorname{span}\{|00\rangle,|11\rangle\}$. On that even subspace, $\Delta^{(3)}(z)$ acts as $\rho_r(z)^2(X\otimes X)=\sigma(r)(X\otimes X)$.

When $r=V$, all four half-edges carry $V$. Both $\Delta^{(3)}(x)$ and $\Delta^{(3)}(y)$ act as $Z^{\otimes4}$, so their common eigenvalue-one subspace consists of the eight basis vectors of even total parity. Evaluating $\Delta^{(3)}(z)$ in the prescribed cyclic order and then regrouping the endpoint factors as $|i_1i_2;j_1j_2\rangle\in\mathcal L_V\otimes\mathcal F_{VV}$ gives the exchanges
\begin{equation}
 \begin{array}{c|c}
 \text{paired basis vectors}&\text{exchange sign}\\ \hline
 |00;00\rangle,\ |11;11\rangle&+1\\
 |00;11\rangle,\ |11;00\rangle&-1\\
 |01;01\rangle,\ |10;10\rangle&+1\\
 |01;10\rangle,\ |10;01\rangle&+1
 \end{array}.
 \label{eq:h8-even-exchanges}
\end{equation}
Selecting the $+1$ eigenvector of $\Delta^{(3)}(z)$ from each pair yields an orthonormal basis for $\operatorname{ran}P_{V;VV}$,
\begin{align}
 \zeta_1&=\frac{|00;00\rangle+|11;11\rangle}{\sqrt2},&
 \zeta_2&=\frac{|00;11\rangle-|11;00\rangle}{\sqrt2},\nonumber\\
 \zeta_3&=\frac{|01;01\rangle+|10;10\rangle}{\sqrt2},&
 \zeta_4&=\frac{|01;10\rangle+|10;01\rangle}{\sqrt2}.
 \label{eq:h8-four-invariants}
\end{align}

Now let $A_b$ be any single-loop operator on $b$ such that $A_b\otimes I_{\{a,C,D\}}$ preserves $\operatorname{ran}P_\Gamma$. For fixed labels $(r_a,r_b,V,V)$, the projector $P_\Gamma$ factors across the two vertices as $P_{r_a;VV}\otimes P_{r_b;VV}$. Setting $r_a=r_b$ and tensoring any vector in $\operatorname{ran}P_{r_b;VV}$ at $v$ with a fixed nonzero vector in $\operatorname{ran}P_{r_a;VV}$ at $u$ produces a valid physical state in $\operatorname{ran}P_\Gamma$. Because different output labels $r'_b$ are orthogonal, each block $(A_b)_{r',r}:\mathcal L_r\to\mathcal L_{r'}$ must satisfy $(1-P_{r';VV})\bigl((A_b)_{r',r}\otimes I_{VV}\bigr)P_{r;VV}=0$.

For $r\in\{0,\eta,s,\bar s\}$ and $r'=V$, the block $(A_b)_{V,r}$ is a vector $w=\sum_{i_1,i_2}w_{i_1i_2}|i_1i_2\rangle\in\mathcal L_V$, and preservation requires $w\otimes\phi_{\sigma(r)}\in\operatorname{span}\{\zeta_1,\zeta_2,\zeta_3,\zeta_4\}$. Because $\phi_{\sigma(r)}$ has even parity, the odd loop components $w_{01}$ and $w_{10}$ have no overlap with $\zeta_3,\zeta_4$ and must vanish. Matching the coefficients of $|00;00\rangle$ and $|11;11\rangle$ to $\zeta_1$ gives $w_{00}=\sigma(r)w_{11}$; matching $|00;11\rangle$ and $|11;00\rangle$ to $\zeta_2$ gives $\sigma(r)w_{00}=-w_{11}$. Since $\sigma(r)^2=1$, these conditions imply $w=0$.

Conversely, for $r=V$ and $r'\in\{0,\eta,s,\bar s\}$, the block $(A_b)_{r',V}$ is a linear functional $\alpha:\mathcal L_V\to\mathbb C$ that must map each $\zeta_k$ into $\mathbb C\phi_{\sigma(r')}$. Applying $\alpha\otimes I_{VV}$ to $\zeta_1$ and $\zeta_2$ gives $\alpha(00)=\sigma(r')\alpha(11)$ and $-\alpha(11)=\sigma(r')\alpha(00)$, forcing $\alpha(00)=\alpha(11)=0$; applying $\alpha\otimes I_{VV}$ to $\zeta_3$ and $\zeta_4$ forces $\alpha(01)=\alpha(10)=0$. Thus both $(A_b)_{V,r}$ and $(A_b)_{r,V}$ vanish, with the two directions checked separately. Under the reducing physical compression, this block separation yields \eqref{eq:hopf-row-obstruction} for the full star algebra $\mathcal A(R_2)$.

\bibliographystyle{hep-references}
\bibliography{references}
\end{document}